\documentclass[a4paper]{article}
\usepackage{epsfig}
\usepackage{graphicx}  
\usepackage{color}     
\usepackage{comment}
\usepackage{color}
\usepackage{lscape}
\usepackage{amssymb}
\date{\today}
 \normalsize

\newcommand{\bmat}{\left(\begin{array}}
\newcommand{\emat}{\end{array}\right)}
\newcommand{\be}{\begin{equation}}
\newcommand{\ee}{\end{equation}}
\newcommand{\bea}{\begin{eqnarray}}
\newcommand{\eea}{\end{eqnarray}}

\def\gtwid{\mathrel{\raise.3ex\hbox{$>$\kern-.75em\lower1ex\hbox{$\sim$}}}}
\def\ltwid{\mathrel{\raise.3ex\hbox{$<$\kern-.75em\lower1ex\hbox{$\sim$}}}}
\def\gev{{\rm \, Ge\kern-0.125em V}}
\def\tev{{\rm \, Te\kern-0.125em V}}

\def    \be            {\begin{equation}}
\def    \ee            {\end{equation}}
\def    \bea           {\begin{eqnarray}}
\def    \eea           {\end{eqnarray}}
\def\eps{\epsilon}
\def\a{\alpha}
\def\b{\beta}

\def\d{\delta}
\def\n{\nu}

\def\th{\theta}

\def\m{\mu}
\def\nn{\nonumber}
\def\d{\delta}
\def\D{\Delta}
\def\s{\sigma}
\def\r{\rho}
\def\t{\theta}

\begin{document}
\renewcommand{\thefootnote}{\fnsymbol{footnote}}

\vspace{.3cm}

\title{\Large\bf Rotated $\mu$\,--\,$\tau$ Symmetry for One Generic Neutrino Mixing Angle: An  analytical Study
}

\author
{ \it \bf  E. I. Lashin$^{1,2}$\thanks{slashin@zewailcity.edu.eg},   N.
Chamoun$^{3,4}$\thanks{nchamoun@th.physik.uni-bonn.de}, C. Hamzaoui$^{5}$\thanks{hamzaoui.cherif@uqam.ca}
 and
S. Nasri$^{6,7}$\thanks{snasri@uaeu.ac.ae}
\\
\small$^1$ Department of Physics, Faculty of Science, Ain Shams University, Cairo 11566,
Egypt.\\
\small$^2$ Centre for Fundamental Physics, Zewail City of Science and
Technology,\\
\small Sheikh Zayed, 6 October City, Giza 12588, Egypt.\\
\small$^3$  Physics Department, HIAST, P.O.Box 31983, Damascus,
Syria.\\
 \small$^4$  Physikalisches Institut der Universit$\ddot{a}$t Bonn, Nu$\ss$alle 12, D-53115 Bonn, Germany.\\
\small$^5$ Groupe de Physique Th\'eorique des Particules,
D\'epartement des Sciences de la Terre et de L'Atmosph\`ere, \\
\small Universit\'e du Qu\'ebec \`a Montr\'eal, Case Postale 8888,  Succ. Centre-Ville,
 Montr\'eal, Qu\'ebec, Canada, H3C 3P8. \\
\small$^6$ Department of Physics, UAE University, P.O.Box 17551, Al-Ain, United Arab Emirates.\\
\small$^7$ Laboratoire de Physique Th\'eorique, ES-SENIA University, DZ-31000 Oran, Algeria.\\
}
\maketitle

\begin{center}
\small{\bf Abstract}\\[3mm]
\end{center}
We find a realization of the $Z_2$-symmetry in the neutrino mass matrix which expresses a rotation of the  $\mu-\tau$ symmetry and is able to impose a generic smallest mixing angle, in contrast to a zero-value
predicted by the usual non-rotated form of the $\mu-\tau$ symmetry. We extend this symmetry for the lepton sector within type-I seesaw scenario, and show it can accommodate the mixing angles, the mass hierarchies and the lepton asymmetry in the universe. We then study the effects of perturbing the specific form of the neutrino mass matrix imposed by the symmetry and compute the resulting mixing and mass spectrum. We trace back this ``low-scale'' perturbation to a ``high-scale'' perturbation, and find realizations of this latter
one arising from exact symmetries with an enriched matter content.
\begin{minipage}[h]{14.0cm}
\end{minipage}
\vskip 0.3cm \hrule \vskip 0.5cm
\section{Introduction}
Flavor symmetry is presumably behind the observed pattern of lepton flavor mixing. The benefits of flavor symmetry are not only limited to deciphering the true origin of
neutrino masses and flavor structures, but it is also used to predict the nine free parameters of the light Majorana neutrino mass matrix  $M_\n$. These nine free parameters comprise the three masses ($m_1, m_2$ and $m_3$), the three mixing angles ($\t_x, \t_y$ and $\t_z$) (commonly known as $\t_{12}, \t_{23}$ and $\t_{13}$),
the two Majorana-type phases ($\r$ and $\s$) and the Dirac-type phase ($\d$). One can set the symmetry at the Lagrangian level which would lead to specific textures of $M_\n$
that one can test whether or not they can accommodate the experimental data summarized in Table \ref{fits}
\begin{table}[h]
\caption{The best-fit values, together with the 1$\s$ and 3$\s$ intervals, for the three mixing angles taken from a global analysis of current experimental data\cite{fog2014}.  (NH, IH) denote respectively Normal and Inverted Hierarchies.}
\centering
\begin{tabular}{cccc}
\hline
\hline
Parameter & Best fit & 1$\sigma$ range & 3$\sigma$ range \\
\hline
\hline
$\sin^2 \theta_{x}/10^{-1}$  \mbox{(NH,IH)} & 3.08 &  2.91 -- 3.25  &  2.59 -- 3.59 \\
\hline
$\sin^2 \theta_{z}/10^{-2}$  \mbox{(NH)} & 2.34 &  2.15 -- 2.54 & 1.76 -- 2.95  \\
$\sin^2 \theta_{z}/10^{-2}$  \mbox{(IH)} & 2.40 &   2.18 -- 2.59 & 1.78 -- 2.98  \\
\hline
$\sin^2 \theta_{y}/10^{-1}$ \mbox{(NH)} & 4.37 & 4.14 -- 4.70 & 3.74 -- 6.26 \\
$\sin^2 \theta_{y}/10^{-1}$  \mbox{(IH)} & 4.55 &  4.24 -- 5.94 &  3.80 -- 6.41
 \\
\hline
\end{tabular}
\label{fits}
\end{table}

A horizontal (flavor) symmetry for leptons was shown to be $S_4$ in \cite{referee1}, and a horizontal symmetry for neutrino mixing was proposed in \cite{referee2}
where the two independent $Z_2$ symmetries to be respected by $M_\n$  were stated. The $Z_2^2$ symmetry which characterizes the phenomenologically successful tripartite model was derived
in a simple way in \cite{lcm}. The works of \cite{referee3} explored the phenomenological consequences of a hidden residual $Z_2$ symmetry in the neutrino mass matrix, whereas a full
flavor symmetry, using the reconstructed residual symmetry, was studied in \cite{referee4}

The $\mu-\tau$ symmetry \cite{Fukuyuma} is treated in many common mixing patterns \cite{literature}
such as tri-bimaximal mixing  \cite{tbm}, bimaximal mixing  \cite{bm} and scenarios of $A_5$ mixing \cite{a5}. This symmetry is determined by fixing one of the two $Z_2$'s which
$M_\n$ respects to reflect exchange between the second and third families.

In \cite{LCHN1,LCHN2}, we realized the $\mu-\tau$ symmetry by two textures, both of which led to a vanishing $\t_z$ angle, and we extended the symmetry to the lepton sector showing it can accommodate the lepton mass hierarchies. However, in order to agree with a non-vanishing experimental value of $\t_z$ we had to resort to ``perturbing'' the texture imposed by
the symmetry. We studied possible forms of perturbations on $M_\n$ originated from perturbations on the Dirac neutrino mass matrix $M_D$ at the high scale within type-I seesaw scenario, and
showed how one can get by this way an experimentally acceptable value for $\t_z$.

However, the exact $\mu-\tau$ symmetry implies $\t_z =0$, so had the experimental value of $\t_z$  not been small enough $(\simeq 9^o)$, one would not have been able to call
 in the perturbative approach onto the $\mu-\tau$ symmetry in order to move $\t_z$ from zero to its large value. Nonetheless, this does not mean that an exact symmetry related to the
  $\mu-\tau$ symmetry can not accommodate the new large value, and the objective of this paper is just to find and investigate such a new symmetry. We shall show that a symmetry ``conjugate''
  to the $\mu-\tau$ symmetry, in the sense that both symmetries represent rotations by the same angle ($\pi$) and so belong to the same conjugacy equivalence class in the group of rotations, leads without perturbation to $\t_z$ equal to an arbitrary value given in advance, and we shall study the consequences of this new ``rotated'' symmetry, called henceforth $S^z$ in contrast to the ``non-rotated'' symmetry $S$ leading to $\t_z=0$. This issue is important since it shows how to move from an exact $\mu-\tau$ symmetry forcing a vanishing value for $\t_z$, that one needs to perturb in order to agree with data, to an equivalent symmetry that agrees with data without a need for perturbations.

 After we find $S^z$, we extend it into the lepton sector and study the resulting mass hierarchies. As we shall see, $S^z$ predicts, in addition to $\t_z$, the value
 $\t_y=\frac{\pi}{4}$, and thus one can study perturbations on $S^z$ in order to account for $\t_y$ slightly different from the acceptable value $\frac{\pi}{4}$. We carry
 out an analytical analysis of perturbing the $M_\n$ form imposed by $S^z$ and compute the ``perturbed'' angles and masses in terms of the perturbation parameters.
 Moreover, we trace back, within type-I seesaw scenarios, this perturbation on $M_\n$ to one affecting $M_D$ at the high scale and express the resulting spectrum in terms of the high-scale
 perturbation. As is considered probable \cite{RGrunning}, we neglect renormalization group effects when running between the two scales, and assume they will not affect the symmetry.
 In \cite{LCHN2}, we perturbed the symmetry $S$ and showed that considerable areas in the parameter space exist accommodating the data with many correlations between the $M_\n$ parameters.
 We expect the same in the case of perturbing $S^z$ which is related ``smoothly'' to $S$. However, this needs to be confirmed by carrying out a complete numerical study scanning the perturbation parameters of $M_D$, which we intend to do in a future work.

 In line with \cite{LCHN2}, the form of perturbation can be generated by assuming exact symmetries at the Lagrangian level, some of which are broken spontaneously by adding new matter fields, and we carry out the ``non-trivial'' work of finding this realization in the case of our rotated symmetry $S^z$.

The plan of the paper is as follows. In Section 2, we review the basic notation for the neutrino mass matrix. In Section 3, we find the new symmetry $S^z$, compute the corresponding mixing and phase angles and the neutrino eigen masses and illustrate the geometrical link between $S$ and $S^z$. In Section 4, we implement $S^z$ into a type-I seesaw scenario. We address the charged lepton sector in Subsection 4.1, whereas we study the neutrino mass hierarchies in Subsection 4.2, and in Subsection 4.3, we comment on leptogenesis induced by $S^z$. In  Section 5, we study deviations on $M_\n$ caused by breaking $S^z$
 in $M_D$, and find the effects on the angles and masses.  In Section 6 we present
a theoretical realization of the perturbed texture. We end by discussion and summary in Section 7. Technical details are reported in three appendices.

\section{Notations}
There are 3 lepton families in the Standard
Model (SM).
The charged-lepton mass matrix relating left-handed (LH) and
right-handed (RH) components is arbitrary, but can always be
diagonalized via a bi-unitary transformation:
\begin{equation}
(V^{l*}_L)^\dagger\; {M}_l\; V^l_R =  \pmatrix{m_e & 0 & 0 \cr 0 & m_\mu & 0 \cr 0 & 0 &
m_\tau} .
\label{diagML}
\end{equation}
In the same manner, we can diagonalize the symmetric Majorana neutrino mass matrix by just one unitary transformation:
\be
V^{\n \dagger} M_\nu\; V^{\n *} \; = \; \left (\matrix{ m_1 & 0 & 0 \cr 0 & m_2 & 0
\cr 0 & 0 & m_3 \cr} \right ), \;
\label{diagM}
\ee
with $m_i$ (for $i=1,2,3$) real and positive.

The mismatch between
$V^l$ and $V^\nu$ leads to the
observed neutrino mixing matrix \footnote{We adopt here (Eqs. \ref{diagML}, \ref{diagM} and \ref{PMNS}) the conventions  used in many works such as those in \cite{conventions1}, where it was shown, as in \cite{conventions2}, that the charged current in a decay process of the form $A \rightarrow B+ \a^+ +\n_\a$ ($\a,\b =e,\mu,\tau$) is given by $J^\r_{cc}=\sum_\a \bar{\n}_\a \gamma_\r \a_L$,  and that a matrix element
$\langle \n_k \b^+ | J^\r_{cc} | 0 \rangle\, J_\r^{A-B}$ ($k=1,2,3$), with $J_\r^{AB}$ denoting the current describing the $A\rightarrow B$ transition, would enter into the relation between the one particle state of neutrino gauge state $|\n_\b \rangle$ and that of the mass eigenstate $|\n_k \rangle$,
and will pick up the combination $(V^l_L)^\dagger_{\b \a} V^\n_{\a k}= (V^{l\dagger}_L V^\n)_{\b k}$ such that $|\n_\b \rangle \propto (V_{\mbox{\tiny PMNS}})_{\b k} |\n_k \rangle$
.}
\begin{eqnarray}
V_{\mbox{\tiny PMNS}} &=& (V^l_L)^\dagger\; V^\nu  .
\label{PMNS}
\end{eqnarray}
In the ``flavor" basis, we have $V^l_L ={\bf 1}$ (the unity matrix) meaning that the charged lepton mass eigen states are the same as the current (gauge) eigen states. We assume that
we are working in this basis implying that the
measured mixing results only from neutrinos $V_{\mbox{\tiny PMNS}} = V^{\nu}$. We justify this by noting that the deviations from $V^l_L \neq {\bf 1}$ are of order
of the ratios of the hierarchical charged lepton masses which are small.

We shall adopt the parametrization of \cite{Xing}, related to other ones by simple relations \cite{LC-zeros}, where the $V_{\mbox{\tiny PMNS}}$
is given in terms of three mixing
angles $(\theta_{x}, \theta_{y}, \theta_{z})$ and three phases ($\delta,\rho,\sigma$),  as follows.
\bea
P_{\mbox{\tiny PMNS}} &=& \mbox{diag}\left(e^{i\rho},e^{i\sigma},1\right)\,, \nn\\
U_{\mbox{\tiny PMNS}} \; &=& \nn R_{y}\left(\t_{y}\right)\; R_{z}\left(\t_{z}\right)\; \mbox{diag}\left(1,e^{-i\d},1\right)\; R_{x}\left(\t_{x}\right)\, \\ &=& \;
\left ( \matrix{ c_{x}\, c_{z} & s_{x}\, c_{z} & s_{z} \cr - c_{x}\, s_{y}
\,s_{z} - s_{x}\, c_{y}\, e^{-i\delta} & - s_{x}\, s_{y}\, s_{z} + c_{x}\, c_{y}\, e^{-i\delta}
& s_{y}\, c_{z}\, \cr - c_{x}\, c_{y}\, s_{z} + s_{x}\, s_{y}\, e^{-i\delta} & - s_{x}\, c_{y}\, s_{z}
- c_{x}\, s_{y}\, e^{-i\delta} & c_{y}\, c_{z} \cr } \right ) \; ,\nn\\
V_{\mbox{\tiny PMNS}} &=& U_{\mbox{\tiny PMNS}}\;P_{\mbox{\tiny PMNS}}\, = \left ( \matrix{ c_{x}\, c_{z} e^{i\rho} & s_{x}\, c_{z} e^{i\sigma}& s_{z} \cr (- c_{x}\, s_{y}
\,s_{z} - s_{x}\, c_{y}\, e^{-i\delta}) e^{i\rho} & (- s_{x}\, s_{y}\, s_{z} + c_{x}\, c_{y}\, e^{-i\delta})e^{i\sigma}
& s_{y}\, c_{z}\, \cr (- c_{x}\, c_{y}\, s_{z} + s_{x}\, s_{y}\, e^{-i\delta})e^{i\rho} & (- s_{x}\, c_{y}\, s_{z}
- c_{x}\, s_{y}\, e^{-i\delta})e^{i\sigma} & c_{y}\, c_{z} \cr } \right ),
\label{defv}
\eea
where $R_{i}\left(\t_{i}\right) (i=x,y,z)$ is the rotation matrix around the $(i-1)^{th}$-axis ($x=1,y=2,z=3 \mbox{ or } 0$) by angle $\t_{i}$, and $s_{i} \equiv \sin\theta_{i}, c_{i} \equiv \cos\theta_{i}$ (later $t_{i} \equiv \tan\theta_{i}$). Note that in this adopted parametrization, the
third column of $V_{\mbox{\tiny PMNS}}$ is real which would be essential later to extract the parameters from the diagonalizing matrix. We write down in Appendix (A) the
elements of the neutrino mass matrix  in the flavor basis and in the adopted parametrization Eq. (\ref{melements}). This helps in viewing directly at the level of the mass matrix that the effect of swapping the indices $2$ and $3$ corresponds to the transformation $\t_{y} \rightarrow \frac{\pi}{2} -
\t_{y}$ and $\d \rightarrow \d \pm \pi$. Hence, for a texture satisfying the $\mu$-$\tau$ symmetry, one can check the correctness of any obtained formula by requesting it to be invariant under the above transformation.

\section{The rotated $S^z$ versus the Non-rotated $S$ Symmetries}
\subsection{The $S$ Symmetry}
The $\mu-\tau$ symmetry is presented by \cite{tbm} at the level of the diagonalizing matrix $V^\n$ as:
\bea \label{VSdef} |V^\n_{\mu i}| &=& |V^\n_{\tau i}|,\;\;\; i=1,2,3\eea which is verified experimentally to a mild degree. Equivalently, we defined in
\cite{LCHN2} the $S$ symmetry incorporating the $\mu-\tau$ symmetry, and which leads to mixing angles in the first quadrant, by an orthogonal real matrix:
\begin{equation}
S =\left(
\begin{array}{ccc}
-1 & 0 & 0 \\
0 & 0 & 1 \\
0 & 1 & 0
\end{array}
\right)
\label{splus}
\end{equation}
Another symmetry is possible by taking the $(1,1)$$^{\mbox{th}}$ entry in $S$ equal to $1$ instead of $-1$, but it led to some mixing angles not lying in the first quadrant, so to fix the ideas, let us just restrict our study to $S$ above. Requiring that $M_\n$ is invariant under $S$:
\begin{eqnarray}\label{FI}
S^{T}\;M_{\nu}\;S = M_{\nu}
\end{eqnarray}
implies that $M_{\nu}$ has a specific form:
\bea
M_{\nu} &=&
\left(
\begin{array}{ccc}
A_{\nu} & B_{\nu} &  -B_{\nu} \\
B_{\nu} & C_{\nu}  & D_{\nu} \\
-B_{\nu} & D_{\nu} & C_{\nu}
\end{array}
\right)
\label{Mn+}
\eea
and that one can diagonalize simultaneously all the matrices $S$, $M_\n$ and $M^*_{\nu} M_{\nu}$ by a unitary matrix $U$ of the form:
  \bea
 \label{U}
U =   \left(
\begin{array}{ccc}
 c_\varphi & s_\varphi &  0 \\
- \displaystyle{{s_\varphi\over \sqrt{2}}}\,e^{-i\,\xi}   & \displaystyle{{c_\varphi\over \sqrt{2}}}\,e^{-i\,\xi} & \displaystyle{{1\over \sqrt{2}}}\\
 \displaystyle{{s_\varphi\over \sqrt{2}}}\,e^{-i\,\xi} & -\displaystyle{{c_\varphi\over\sqrt{2}}}\,e^{-i\,\xi} & \displaystyle{{1\over \sqrt{2}}} \end{array}
 \right),
\eea
such that
\bea
 U^{T}\;M_{\nu}\; U = M_\n^{\mbox{\tiny Diag}}= \mbox{Diag}
 \left(
 M_{\n\,11}^{\mbox{\tiny Diag}},\; M_{\n\,22}^{\mbox{\tiny Diag}},\; M_{\n\,33}^{\mbox{\tiny Diag}} \right)&,&
 U^{\dagger}\;M^*_{\nu}\;M_{\nu}\; U
 = \mbox{Diag} \left( m_1^2,m_2^2,m_3^2 \right)
 \label{diag}
\eea

The angles $\varphi$ and $\xi$ are determined by the requirement that $U^\dagger M^*_{\nu} M_{\nu} U$ is a diagonal matrix and found to be equal to
\bea
\label{varphi} \tan \left(2 \varphi\right) =
\frac{2 \, \sqrt{2} \, \left|b_\n\right| }
{c_{\nu} - a_{\nu} - d_{\nu} }, && \xi = \mbox{Arg}\left(b_\n\right),
\label{vars}
\eea
where we write the squared mass matrix in the form:
\bea
\label{M*MS+}
 M_{\nu}^* \,M_{\nu} &= &
 \left(
 \begin{array}{ccc}
 a_\n & b_\n & - b_\n \\
b_\n^* &  c_\n & d_\n\\
-b_\n^* & d_\n & c_\n
\end{array}
\right)
\label{Mn*Mn+},
\eea
The relations between the entries of $M_{\nu}^* \,M_{\nu}$ and those of $M_\n$ are written in Eq. (\ref{abs+}) in Appendix A. Also, we write there the mass spectrum corresponding to
 the eigenvalues of $M_\n$ and $M_{\nu}^* \,M_{\nu}$ in Eq. (\ref{Mdiags+}) and  Eq. (\ref{msqs+}) respectively. Multiplying $U$ by a diagonal phase matrix
\bea
Q=\mbox{Diag} \left\{
\exp{\left[{-\displaystyle{\frac{i}{2}} \mbox{Arg} \left(M_{\n\,11}^{\mbox{\tiny Diag}}\right)}\right]},\;
\exp{\left[{-\displaystyle{\frac{i}{2}} \mbox{Arg} \left(M_{\n\,22}^{\mbox{\tiny Diag}}\right)}\right]},\;
\exp{\left[{-\displaystyle{\frac{i}{2}} \mbox{Arg} \left(M_{\n\,33}^{\mbox{\tiny Diag}}\right)}\right]}
\right\},
\label{Q}
\eea
in order to get rid of the phases of  $M_\n^{\mbox{\tiny Diag}}$, and rephasing the charged lepton
fields so that to make the conjugate of $UQ$ in the same form as the adopted parametrization of $V_{\mbox{\tiny PMNS}}$ (a real third column) we find, in addition to $\t_z=0$, the following mixing and phase angles:
 \bea
 \label{mixingS+}
 &&\t_{y}= \pi / 4,\;\; \t_{x}=\varphi ,\;\; \nn \\
 && \rho = {1\over 2}\, \mbox{Arg}\left(M_{\n\,11}^{\mbox{\tiny Diag}}\, M_{\n\,33}^{*\mbox{\tiny Diag}}\right),\;\;
 \sigma = {1\over 2}\,\mbox{Arg}\left(M_{\n\,22}^{\mbox{\tiny Diag}}\,M_{\n\,33}^{*\mbox{\tiny Diag}}\right),\;\; \delta = 2\pi - \xi.
 \eea
Thus the $\mu-\tau$ symmetry $S$, which is diagonalized, as well as $M_\n$, by $U=R_y(\pi/4)\, R_z(0)\, X$, where
$X=\mbox{diag}\left(1,e^{-i\d},1\right)\; R_{x}\left(\t_{x}\right)\,\mbox{diag}\left(e^{i\rho},e^{i\sigma},1\right)\,$ involves the phase angles and the rotation $R_x$, fixes two mixing angles $\left(\t_{y} = \pi/4,\, \t_z=0\right)$ whereas the other parameters are determined, without fine tuning, by $M_\n$.

\subsection{The $S^z$-Symmetry}
 Now we see how one should proceed in order to seek the new symmetry $S^z$ leading directly to a non-vanishing value $\t_z$, in that it should be diagonalized by
 $U^z=R_y(\pi/4) R_z(\t_z) X$. The matrix $U^z$ can be written in a form  related to $U$ as,
 \bea
 U^z &=& W U\label{Uz},
\eea
provided $W$ is given by the following rotation matrix
\bea
W &=& R_y(\pi/4) R_z(\t_z) R_y^{-1}(\pi/4) =
\left(
\begin{array}{ccc}
 c_z & \displaystyle{{s_z\over \sqrt{2}}} &  \displaystyle{{s_z\over \sqrt{2}}} \\
-\displaystyle{{s_z\over \sqrt{2}}} &  \displaystyle{c^2_{(z/2)}} &  -\displaystyle{s^2_{(z/2)}}\\
-\displaystyle{{s_z\over \sqrt{2}}} & -\displaystyle{s^2_{(z/2)}} & \displaystyle{c^2_{(z/2)}}
\end{array}
\right).
\label{W}
\eea
Starting from Eq. (\ref{diag}), we find that $U^z$ diagonalizes simultaneously a new neutrino mass matrix $M_\n^z$ and its hermitian square $M_\n^{*z} M_\n^z$ as,
\bea
U^{zT}\,M^z_\n\, U^z &=&  M_\n^{\mbox{\tiny Diag}},\nn\\
U^{z\dagger}\,M_\n^{*z}\, M^z_\n\, U^z &=&  M_\n^{\mbox{{\tiny Diag}}*}\, M_\n^{\mbox{\tiny Diag}},
\label{diagz}
\eea
where the new ``rotated'' neutrino mass matrix $M_\n^z$ is related to $M_\n$ by a similarity transformation given by the rotation $W$:
\bea
M^z_\n &=& W\, M_\n\, W^T
\label{Mz}.
\eea
We notice that the mass spectrum is the same for $M_\n$ and $M_\n^z$ expressing the fact that the eigenmasses are invariant under a change of basis.

The ``rotated'' symmetry $S^z$ can equally be defined as a similarity transformation applied to $S$ via the rotation matrix $W$:
\bea
\label{Sz}
S^z &=& W \, S\,  W^T =
\left(
\begin{array}{ccc}
 -c_{2z} & \sqrt{2}\, s_z\, c_z &  \sqrt{2}\, s_z\, c_z \\
\sqrt{2}\, s_z\, c_z &  -s_z^2 &  c_z^2\\
\sqrt{2}\, s_z\, c_z & c_z^2 & -s_z^2
\end{array}
\right)
\eea
since we see that  $M^z_\n$ is form invariant under $S^z$:
\bea
S^{zT} \, M_\n^z \, S^z &=& M_\n^z.
\label{FIz}
\eea
We stress here that $M_\n^z$ is the `physical' neutrino mass matrix in our setup which predicts $\theta_z \ne 0$. The invariance of $M^z_\n$ under the symmetry $S^z$ implies the following form:
\bea
\label{Mnz}
M^z_{\n} =
\left(
\begin{array}{ccc}
A^z_\n & B^z_\n &  B'^z_\n \\
B_\n^z & C_\n^z  & D_\n^z \\
B'^z_\n & D^z_\n & C'^z_\n
\end{array}
\right),
\eea
where $A_\n^z$, $B_\n^z$, $C_\n^z$ and $D_\n^z$ are arbitrary independent complex parameters, while $B'^z_\n$ and
$C'^z_\n$ are dependent and given as
\bea
B'^z_\n &=& -\left(1-2\,t_z^2\right) B^z_\n -\sqrt{2}\, t_z\, \left(A^z_\n - C^z_\n-D^z_\n\right),\nn  \\
C'^z_\n &=& \left(1-2\,t_z^2\right) C^z_\n + 2\, t_z^2\, \left(A^z_\n - D^z_\n\right) + 2\, \sqrt{2}\, t_z\, \left(1-t_z^2\right)\, B^z_\n.
\label{Mzncons1}
\eea

We state in Eq. (\ref{relAnzAn}) in Appendix A the expressions of the $M_\n^z$ elements (Eq. (\ref{Mnz})) in terms of the ``non-rotated'' $M_\n$ elements (Eq. \ref{Mn+}).
In addition, Eq. (\ref{mz2Par}) there expresses the general form of the ``rotated'' square mass matrix $M_\n^{z*} M_\n^z$  in terms of the ``non-rotated'' parameters of the square
 mass  $M_\n^{*} M_\n$ defined in Eq. (\ref{abs+} ). Having established the relations between the rotated $\left\{M_\n^z,\; M_\n^{z*} M_\n^z\right\}$ and the non-rotated $\left\{M_\n,\; M_\n^{*} M_\n\right\}$ mass matrices, we stress again that the mass spectra corresponding to both sets of $\left\{M_\n^z,\; M_\n^{z*} M_\n^z\right\}$ and $\left\{M_\n,\; M_\n^{*} M_\n\right\}$  are the same, reflecting the fact that the eigenmasses are invariant under a change of basis.

Plugging Eqs. (\ref{U}, \ref{W}) in Eq. (\ref{Uz}) gives the most general unitary matrix diagonalizing the commuting matrices  $S^z$, $M^z_\n$ and $M_\n^{z*} M^z_\n$.
Alternatively one can start by assuming the form of $S^z$ given in Eq. (\ref{Sz}), ignoring any relation between $S^z$ and $S$, and deduce from it
the mixing and phase angles. Actually, one
 assumes the invariance given in Eq. (\ref{FIz}), which implies that $ S^z$ commutes with  $M_{\nu}^z$ and $M_{\nu}^{z*}$ and consequently also with $M_{\nu}^{z*} M_{\nu}^z$ and $M_{\nu}^z M_{\nu}^{z*}$, then the diagonalizing matrices of $S^z$ include the diagonalizing matrix $U^z$ present in Eq. (\ref{diagz}). The eigenvalues of $S^z$ are: $\left\{- 1,- 1, 1\right\}$ corresponding respectively to the normalized eigen vectors,
\be
 v_1= \left(\,0,\, -\displaystyle{{1\over \sqrt{2}}}, \displaystyle{{1\over \sqrt{2}}}\,\right)^{\mbox{\tiny T}}, \,
 v_2= \left(\,c_z,\,-\displaystyle{{s_z\over\sqrt{2}}},\,\displaystyle{{s_z\over\sqrt{2}}}\,\right)^{\mbox{\tiny T}}, \,
 v_3=\left(\,s_z, \displaystyle{{c_z\over\sqrt{2}}},\,\displaystyle{{c_z\over\sqrt{2}}}\,\right)^{\mbox{\tiny T}}.
\label{eigvsz}
\ee
In constructing  the general form (up to a diagonal phase matrix) of the unitary diagonalizing matrix of $S^z$ one should care about the two-fold degenerate eigenvalue $-1$, which entails the freedom for a unitary transformation defined  by an angle $\varphi$ and phase $\xi$ in its eigenspace to get new eigen vectors in the following form:
\be
\overline{v}_1  = s_\varphi\, e^{-i\,\xi}\, v_1 +  c_\varphi\, v_2,\;\;
\overline{v}_2  =  -c_\varphi\, e^{-i\,\xi}\, v_1 + s_\varphi\, v_2,\;\;
\overline{v}_3  = v_3.
\label{reigvsz}
\ee
 The ambiguity in ordering the eigenvectors of $S^z$ is fixed by requiring that the mixing angles, contained in the unitary matrix $U^z$ diagonalizing $S^z$, should fall all in the first quadrant. The desired order  turns out to correspond  $\left\{- 1,- 1, 1\right\}$ and thus the matrix $U^z$ assumes  the following form:
 \bea
 \label{Uzgeneral}
U^{z}=\left[\overline{v}_1, \overline{v}_2, \overline{v}_3\right] =
\left(
\begin{array}{ccc}
 c_z c_\varphi & c_z s_\varphi &  s_z \\
-\displaystyle{{1\over \sqrt{2}}} \left(s_z c_\varphi+ s_\varphi\,e^{-i\,\xi} \right)  & -\displaystyle{{1\over \sqrt{2}}}\,\left(s_z s_\varphi - c_\varphi\,e^{-i\,\xi} \right)& \displaystyle{c_z\over\sqrt{2}}\\
  -\displaystyle{{1\over \sqrt{2}}}\,\left(s_z c_\varphi - s_\varphi\,e^{-i\,\xi}\right) & -\displaystyle{{1\over \sqrt{2}}}\,\left(s_z s_\varphi + c_\varphi\,e^{-i\,\xi} \right) & \displaystyle{c_z\over\sqrt{2}}
  \end{array}
  \right).
\eea
We can check that Eq. (\ref{Uzgeneral}) is consistent with Eqs. (\ref{U}, \ref{Uz} and \ref{W}). The specific form of $U^z$ of Eq. (\ref{Uzgeneral}) which diagonlizes also the hermitian matrix $M_\n^{z*} M^z_\n$, which commutes with $S^z$, leads to the same mixing angle $\varphi$ and phase angle $\xi$ as determined in Eq. (\ref{vars}).

In order to get a positive mass spectrum  we use the freedom of multiplying $U^z$ by a diagonal phase matrix $Q$ which turns out be the same as found in Eq. (\ref{Q}),
\bea
(U^z\; Q)^T\; M_\nu^z\; (U^z\; Q) &=& \mbox{Diag}\left(m_1,\;m_2,\; m_3\right).
\label{UzQ}
\eea
Moreover, we re-phase the charged lepton fields in order to make real the $3^{rd}$ column of the conjugate of $(U^z\; Q)$ in accordance with the adopted parametrization for $V_{\mbox{\tiny{PMNS}}}$ in Eq.(\ref{defv}), so that to identify the mixing and phase angles. We find that the new ``rotated'' $\mu$--$\tau$ symmetry realized through $S^z$ entails the followings:
 \bea
 \label{mixingSz}
 &&\t_y= \pi / 4,\;\; \t_x=\varphi ,\;\; \t_z= \t_z ,\nn \\
 && \rho = {1\over 2}\, \mbox{Arg}\left(M_{\n\,11}^{\mbox{\tiny Diag}}\,M_{\n\,33}^{*\mbox{\tiny Diag}}\right),\;\;
 \sigma = {1\over 2}\,\mbox{Arg}\left(M_{\n\,22}^{\mbox{\tiny Diag}}\,M_{\n\,33}^{*\mbox{\tiny Diag}}\right),\;\; \delta = 2\pi - \xi.
 \eea
Starting with $S^z$ as a symmetry imposed on the neutrino mass matrix and parameterized by an angle $\t$ (put by hand), we get two predictions ($\t_y=\pi /4, \t_z = \t$) which are
are phenomenologically viable especially for predicting a non vanishing $\t_z$, which is at our disposal through defining the symmetry $S^z$. Adjusting $\t_x$
to accommodate the experimental value of $\t_x \simeq 33.7^o$ does not require a special adjustment for the mass parameters $a_\n, b_\n, c_\n, d_\n$. There is still the need for a small deviation from the exact $S^z$ symmetry to shift $\t_y$ to its experimental value $\t_y \simeq 41^o$.

The various neutrino mass hierarchies can also be produced as can be seen from Eq.(\ref{msqs+}) and Eq.(\ref{vars}) where
 the measurable four observables (the three masses and the mixing angle $\varphi = \t_x$) are given in terms of four parameters $a_\n, \left|b_\n\right|, c_\n$, and $d_\n$.
Therefore, one can solve the four given equations to get $a_\n, \left|b_\n\right|, c_\n$, and $d_\n$ in terms of the masses and the angle $\varphi$.

One can also present the new ``rotated'' $\mu-\tau$ symmetry at the level of the diagonalizing matrix $V^\n$ starting with the corresponding definition for the
$\mu-\tau$ symmetry Eq. (\ref{VSdef}) and using Eq. (\ref{Uz}) with $W= R_y\left(\pi/4\right)\, R_z\left(\t_{z0}\right)\, R_y^{-1}\left(\pi/4\right)$  to get the following constraints which can be taken now as the definition of
the ``rotated'' $\mu-\tau$ symmetry.
\bea
\label{VSzdef}
\left|\frac{s_{z0}}{\sqrt{2}} V^\n_{1j}+c^2_{z0/2} V^\n_{2j}-s^2_{z0/2} V^\n_{3j}\right|&=&\left|\frac{s_{z0}}{\sqrt{2}} V^\n_{1j}-s^2_{z0/2} V^\n_{2j}+c^2_{z0/2} V^\n_{3j}\right|, \;\;\; (j=1,2,3)
\eea
Starting from the above defining texture, one finds that it implies $\t_y = \pi / 4, \t_z = \t_{z0}$ whereas $\t_x, \d, \r, \s$ are not constrained. Moreover, we can check
experimentally the approximate validity of Eq. (\ref{VSzdef}) for the best fit values ($\t_x=33.7^o,\t_y = 41.9^o, \t_z=8.85^o$) in that we get:
\be
 \label{check}
\left.
\begin{array}{lll}
j=1 &\Rightarrow& \left|0.00502  - 0.4127 e^{-i\d}\right| \simeq \left|0.0047 - 0.37079 e^{-i\d}\right|,\\
&\mbox{for } \d=0,\;\;\;\;(\d=\pi/2)& 0.40769 \simeq 0.36603,\;\;\;( 0.41273 \simeq 0.37082), \\\\
j=2 &\Rightarrow& \left|0.00335 + 0.6188 e^{-i\d}\right| \simeq \left|0.00317 + 0.5559 e^{-i\d}\right|, \\
&\mbox{for } \d=0,\;\;\;(\d=\pi/2)& 0.62220 \simeq 0.55916,\;\;\;(0.61881\simeq 0.55591),
\\\\
j=3 &\Rightarrow& 0.6683123 \simeq 0.743880.
\end{array}
\right\}
\ee
The approximate equalities, which are nearly respected in Eq.~(\ref{check}), suggest that the $S^z$ symmetry  is not ruled out by data, but rather is a viable one, and the small deviations can be attributed to small symmetry breaking effects. Moreover, the numerical values of the discrepancies between data and predictions depend on our choice of the best fit values, and can be smaller
for other experimentally acceptable values. Thus, the defining constraints of $S^z$ at the level of $V^\n$ are fulfilled to a satisfactory level. This motivates then the adoption of the $S^z$ symmetry as a starting point towards imposing the form of $M_\n$. Furthermore, it is plausible to probe and test  the ``mixing sum rule'' (Eq. \ref{VSzdef}) at the planned future long baseline neutrino experiments.

\subsection{Equivalence class of $S$ and $S^z$}
We show now that both $S$ and $S^z$ belong to the same conjugation class in the group of $SO(3)$ \footnote{We say that two elements $g_1$ and $g_2$ of a group $G$ are conjugate
if there is an element $g \in G$ such that $g_1=gg_2g^{-1}$, and this conjugation is an equivalence relation.}.
In fact, one can write
\bea \label{rotationS}
S &=& R_y\left(\pi/2\right)\,R_z\left(0\right)\,R_x\left(\pi\right),
\eea
which shows that $S$ is a rotation matrix. If we denote a rotation matrix representing a rotation around an axis whose unit vector is ${\bf \hat{n}}$ through angle $\t$
by $R_{{\bf \hat{n}}}(\t)$ then we have, after using Eq. (\ref{W}),
\bea
\label{Sn}
S=R_{{\bf \hat{n}_0}}(\pi) &,& {\bf \hat{n}_0} = (0,1/\sqrt{2}, 1/\sqrt{2})^T,\\
\label{Sznz} S^z=R_{{\bf \hat{n}^z}}(\pi) &,& {\bf \hat{n}^z} =W \,{\bf \hat{n}_0}= (s_z,c_z/\sqrt{2}, c_z/\sqrt{2})^T.
\eea
The following well known result in rotations group theory:
\bea
\label{conjugation}
R_{{\bf \hat{n}'}}(\t) &=& R\, R_{{\bf \hat{n}}}(\t)\; R^{-1},
\eea
where ${\bf \hat{n}'} = R\, {\bf \hat{n}}$ is the result of rotating ${\bf \hat{n}}$ via the rotation $R$, shows
that rotations through the same angle belong to the same conjugation equivalence class, which is a $2$-dimensional surface  (with two parameters determined by the axis ${\bf \hat{n}'}$) in the $3$-dimensional
manifold of rotations. We have now
\bea \label{SSzconjugation}
S^z=R_{{\bf \hat{n}^z}}(\pi)=W\, R_{{\bf \hat{n}_0}}(\pi)\, W^{-1}= W\,S\,W^T
\eea
where $W$ of Eq. (\ref{W}) represents the rotation $R_{\bf \hat{m}}(\t_z)$ through the angle $\t_z$ around the axis defined by
${\bf \hat{m}} \propto {\bf \hat{n}_0} \times {\bf \hat{n}^z} \propto (0,1,-1)^T$ (which is an eigen vector of $W$ with eigenvalue $+1$).

In summary, the symmetry $S^z(\t)$ represents a $1$-parameter path inside the $2$-parameters equivalence class of the $\mu-\tau$ symmetry in the group $SO(3)$. When we impose $S^z$-invariance on the neutrino mass matrix $M_\n^z$ (Eq. \ref{FIz}), we obtain Eq. (\ref{Mz}), and so get the same eigen masses as in the $\mu-\tau$ symmetry case. However, the resulting mixing matrices, as well as the characteristic defining relations between their entries, are different (cf. Eqs. \ref{VSdef} and \ref{VSzdef}), and for the $S^z$ case they are not invariant under
the interchange $\mu \leftrightarrow \tau$. This shows the geometrical link between $S$ and $S^z$ and justifies our calling of $S^z$ as a ``rotated'' $\mu-\tau$ symmetry.

Finally, it is important to realize that any general discrete Abelian flavor symmetry applied to lepton sector  does not necessarily imply zero mixing angle $\th_z$. The present work  demonstrates this assertion by working out a specific case represented by the rotated $\mu-\tau$ symmetry $S^z$. In other previous works, it was shown how to implement a discrete flavor symmetry $(Z_2)^3$ \cite{z23nontr} and a continuous flavor symmetry $U(1)$ \cite{u1nontr} in order to produce non-zero mixing angle $\th_z$. The same facts  apply to
the case of non-Abelian flavor symmetry, as one can find examples of $A_4$ flavor symmetry \cite{king,ding} yielding non-vanishing $\th_z$.

\section{The seesaw mechanism and the $S^z$ symmetry}

We extend now the $S^z$-symmetry in the Lepton sector at the Lagrangian level, then we use the type-I seesaw scenario to treat the effective neutrino mass matrix, with consequences on leptogenesis.

 \subsection{The charged lepton sector}
We start with the part of the SM Lagrangian which gives masses to the charged leptons:
\bea
\label{L1}
\mathcal{L}_1 =
Y_{ij}^z \, \overline{L} _i
\,
\phi \, \ell _j ^c,
\eea
where the SM Higgs field
$\phi$ and the right handed (RH) leptons   $\ell _j ^c$ are assumed to be singlet under $S^z$, whereas the left handed (LH) leptons transform as:
\bea
L_i &\longrightarrow& S^z_{ij}\, L_j.
\eea
Invariance under $S^z$ leads to:
\bea
S^{zT}\, Y^z &=& Y^z,
\eea
and this forces the Yukawa couplings to look like:
\bea
Y^z &=& \left(\begin{array}{ccc} \sqrt{2} t_z a & \sqrt{2} t_z b & \sqrt{2} t_z c \\
a & b & c \\
a & b & c \end{array}\right),
\eea
When $\phi$ gets a vacuum expectation value (vev) $v$, we have:
\bea
M^z_l\, M_l^{z\dagger}
&=&
v^2 \,
\left(
\begin{array}{ccc}
2 t_z^2 & \sqrt{2} t_z & \sqrt{2} t_z \\
\sqrt{2} t_z & 1 & 1 \\
\sqrt{2} t_z & 1 & 1
\end{array}
\right)
\,
\left(|a|^2 + |b|^2 + |c|^2\right).
\eea
The eigenvalues of $M^z_l\, M_l^{z\dagger}$ are $2 v^2 (1+t_z^2) \,\left(|a|^2 + |b|^2 + |c|^2\right)$ (with eigenvector $ \left[\,s_z,\, c_z/\sqrt{2},\, c_z/\sqrt{2}\,\right]^{\mbox{\tiny T}}$)
and $0$ (with an eigenspace spanned by the normalized eigenvectors $\sqrt{1+2t_z^2/(1+t_z^2)} \left[\,1,\,-t_z/\sqrt{2},\,-t_z/\sqrt{2}\,\right]^{\mbox{\tiny T}}$ and $\displaystyle{{1\over \sqrt{2}}} \left[\,0,\,-1,\,1\,\right]^{\mbox{\tiny T}}$) , then the charged lepton mass hierarchy can not be produced, and the nontrivial diagonalizing matrix contradicts our assumption of being in the flavor basis. To remedy this, we need to introduce many SM Higss doublets $\phi_i$ coupled to the lepton LH doublets through the standard renormalizable  Yukawa  interaction terms
\bea
\mathcal{L}_2 = f^z_{ikr} \, \overline{L} _i
\, \phi _k \,  \ell _r ^c.
\eea
Regarding the flavor-changing neutral Yukawa interaction, it could be suppressed by properly adjusting the relevant Yukawa coupling combinations\cite{grim-hag}.

We assume the $\phi_k$'s transform under $S^z$ as:
\bea
\phi _i \longrightarrow S^z_{ij}\, \phi _j.
\eea
Invariance under $S^z$ implies,
\bea
S^{zT} f_r^z S^z= f_r^z,  & \mbox{where }&
\left(f_r^z\right)_{ij}= f^z_{ijr},
\eea
and thus we can show that $f^z_r$ can be parameterized as
\bea
f^z_r =
\left(
\begin{array}{ccc}
-\displaystyle{{1\over \sqrt{2}\,\,t_z}}\, \left(F^r+G^r\right)+\sqrt{2}\, t_z\, K^r + N^r + P^r & F^r+G^r-K^r & K^r \\
F^r & \sqrt{2}\,t_z\,\left(K^r-F^r\right)+P^r & N^r \\
G^r & \sqrt{2}\,t_z\,\left(K^r-G^r\right)+N^r & P^r
\end{array}
\right),
\label{yminih}
\eea
We use this parametrization for $f^z_r$ in a way to put in its third column independent entries, so that when the the Higgs fields $\phi_k^\circ$ acquire vevs
$\left( v_k=\langle\phi_k^\circ\rangle\right)$ and when we assume having a hierarchy of the form:
 $v_1, v_2 \ll v _3$ then
\bea
(M^z_l) _{ir} \simeq
v_3 f^z_{i3r}
&\simeq& v_3
\left(\begin{array}{ccc} K^1 & K^2 & K^3 \\
N^1 & N^2 & N^3 \\
P^1 & P^2 & P^3 \end{array}\right).
\label{ML1}
\eea
Interpreting each row of Yukawa couplings in the mass matrix Eq. (\ref{ML1}) as a complex valued vector having norm defined in the standard way, and assuming that the ratio between
  the moduli of these Yukawa vectors matches the corresponding one between lepton masses  as $\left|K\right| : \left|N\right| : \left|P\right|
\sim m_e : m_\mu : m_\tau$, one can show, as was done in \cite{lcm,LCN-z23u1}, that the LH charged lepton fields needs to be infinitesimally rotated in order to diagonalize the charged lepton mass matrix, which validates our assumption of  working in the flavor basis to a good approximation. In fact, we have verified numerically that the diagonalizing matrix for the charged lepton mass matrix is nearly the identity matrix with off-diagonal elements of the order of the acute hierarchical charged lepton mass ratios in line with \cite{ma}.


On the other hand,
we could have introduced many $SM$-singlet scalar fields $\Delta _k$, in addition to one SM Higss $\phi$, coupled to the lepton LH doublets through the dimension-5 operator
(only one SM-Higgs field is chosen for purposes related to suppressing flavor--changing neutral currents but at the expense of renormalizability):
\bea
\mathcal{L}_2 = \displaystyle \frac{f^z_{ikr}}{\Lambda}
 \, \overline{L} _i
\,
\phi \, \Delta _k \,  \ell _r ^c.
\label{lagdim5}
\eea
The singlet fields $\Delta_k$ can mix with the SM Higgs field, which may affect the collider phenomenology. We shall not discuss this possibility, in particular that the couplings responsible for the mixing of $\Delta_k$ and $\phi$ are, a priori,
independent of the parameters appearing in the Majorana neutrino mass terms, and each set of parameters can be tuned separately to satisfy the phenomenological constraints.

We assume the $\Delta_k$'s transform under $S^z$ as:
\bea
\Delta _i \longrightarrow S^z_{ij}\, \Delta _j.
\eea
Thus the resulting Yukawa couplings $f^z_{ikr}$ in Eq.(\ref{lagdim5}) would follow the same pattern as in Eq. (\ref{yminih}).
When the fields $\Delta _k$ and the neutral component of the Higgs field $\phi^\circ$ acquire vevs
$\left(\langle\Delta _k\rangle = \delta _k,\, v=\langle\phi^\circ\rangle\right)$ and when we assume having a hierarchy of the form:
 $\delta_1, \delta _2 \ll \delta _3$ then
\be
(M^z_l) _{ir} \simeq
\displaystyle \frac{v f^z_{i3r} }{\Lambda} \delta _3.
\label{ML2}
\ee
We can follow now the same procedure, described above when using many SM Higgs doublets $\phi_i$, to get the diagonalized charged leptons while still working in the flavor basis to a good approximation.

\subsection{Neutrino mass hierarchies}

The effective light LH neutrino mass
matrix is generated through the seesaw mechanism formula
\bea
\label{seesaw}
M^z_\n&=&M^z_D\, \left(M_R^{z}\right)^{-1}\, M_D^{z\mbox{\tiny T}},
\eea
The Dirac neutrino mass matrix $M^z_D$, in case of single SM-Higss,
originates from the Yukawa term
\bea
\label{MnD1}
g^z_{ij}\; \overline{L}_i\; i\tau_2\, \phi^* \n_{Rj},
\eea
after the Higgs field gets a vev, whereas the
symmetric Majorana neutrino mass matrix $M^z_R$ comes from a term
\bea
\label{MnR1}
{1\over 2}\, \n_{Ri}^T\, C^{-1}\, \left(M^z_R\right)_{ij}\, \n_{Rj}.
\eea
where $C$ is the charge conjugation matrix.

We assume the RH neutrino to transform under $S^z$ as:
\bea
\n_{Rj } \longrightarrow S^z_{jr}\, \n_{Rr},
\eea
and by $S^z$-invariance we have
\bea
S^{zT}\, g^z\, S^z = g^z &,& S^{zT}\, M^z_R\, S^z = M^z_R.
\eea
Knowing the required textures for the above constraints in the case of $S$ symmetry
\bea \label{MD}
M_D = g v = \,
\left(\begin{array}{ccc}
A_D & B_D & - B_D\\
E_D & C _ D & D _D \\
-E_D & D_D & C_D \end{array}\right) &,&
M_R =
\left(\begin{array}{ccc}
A_R & B_R & - B_R\\
B_R & C _ R & D _R \\
-B_R & D_R & C_R \end{array}\right),
\label{formMRD}
\eea
one can find the necessary textures in the case of $S^z$ symmetry by simply computing
\bea
M^z_D = W\, M_D\, W^T &,& M^z_R = W\, M_R\, W^T
\eea
Whereas the parametrization of the symmetric $M^z_R$ is similar to $M^z_\n$ Eq. (\ref{Mnz}), the symmetry $S^z$ dictates the following form of $M_D^z$    \bea
   \label{MzD}
M^{z}_D &=& \,
\left(\begin{array}{ccc}
A^{z}_D & B^{z}_D & B'^{z}_D\\
E^{z}_D & C^{z}_D & D^{z}_D \\
E'^{z}_D & D'^{z}_D & C'^{z}_D
\end{array}\right),
\eea
where $A^z_D$, $B^z_D$, $C^z_D$, $D_D^z$ and $E_D^z$ are arbitrary independent complex parameters, while $B'^z_D$,
$C'^z_D$, $D'^z_D$ and  $E'^z_D$ are dependent and given as
\bea
B'^{z}_D &=& -B^{z}_D + \sqrt{2} t_z\,\left(D^{z}_D+ C^{z}_D -A^{z}_D\right) + 2 t_z^2 E^{z}_D,  \nn \\
E'^{z}_D &=& -\left(1-2t_z^2\right) E_D^{z} + \sqrt{2} t_z\, \left(D_D^{z}+C_D^{z}-A_D^{z}\right),\nn \\
 D'^{z}_D  &=& D_D^{z} + \sqrt{2} t_z\, \left(E^{z}_D-B^{z}_D\right),\nn  \\
 C'^{z}_D  &=& \left(1-2t_z^2\right) C_D^{z} + \sqrt{2}\, t_z\, B_D^{z} + 2\, t_z^2\, \left(A_D^{z} - D_D^{z}\right) + \sqrt{2}\, t_z\, \left(1-2 t_z^2\right)\, E_D^{z}
 \label{Mzdcon1}.
 \eea

In the Appendix (A), we summarize in useful formulae (Eqs. (\ref{MD1}) upto (\ref{MDz2Par})) all the relevant relations between the entries corresponding to the set of mass matrices $\left\{M_D^z, M_D^{z\dagger}\, M_D^z\right\}$ and those of  $\left\{M_D, M_D^\dagger\, M_D\right\}$.

The seesaw formula implies that
\be
\mbox{det}\left(M_\n^{z*}\,M^z_\n\right) = \mbox{det}\left(M_D^{z\dagger} \,M_D^z\right)^2\; \mbox{det}\left(M_R^{z*}\,M_R^z\right)^{-1}.
\label{relspec}
\ee
and since the determinant, or equivalently the product of eigenvalues, does not change when changing the basis, then the relevant spectra for $M_\n^{z*}\,M^z_\n$, $M_R^{z*}\,M^z_R$ and $M_D^{z\dagger} \,M_D^z$, with the help of Eq. (\ref{msqs+}) and Eq. (\ref{specMD}),  can be written as,
\be
\left\{\; c_{\n,R,D} + d_{\n,R,D},\; {a_{\n,R,D} + c_{\n,R,D} - d_{\n,R,D}\over 2} \pm {1\over 2} \sqrt{\left(a_{\n,R,D}+d_{\n,R,D}-c_{\n,R,D}\right)^2 + 8\, \left|b_{\n,R,D}\right|^2}
\;\right\}.
\label{totspec}
\ee
The mass spectrum and its hierarchy type are determined by the eigenvaules presented in Eq.(\ref{totspec}). As one of  the simplest realizations which can be envisaged from Eq.(\ref{relspec}), one can adjust the spectrum of $M_R^{z*}\,M_R^z$ so that to follow the same kind of hierarchy as $M_\n^{z*}\,M_\n^z$. However, this does not necessitate  that  $M_D^{z\dagger}\,M_D^z$ would behave in the same manner. Also, this is by no means an exclusive example, as there might be other possible realizations producing the same desired hierarchy, and what is mentioned is a mere simple possibility.

Later, we shall need the general forms of the symmetric and general matrices which are ``sign-reversed'', i.e. multiplied by $-1$, under $S^z$. Therefore, for the sake of  completeness and necessity, we state  in  the Appendix (A) all the constraints imposed by such kind of``sign-reversed'' symmetry  (Eq. (\ref{invszsymM}) and Eq. (\ref{invszgenM})).

\subsection{Leptogenesis}

In \cite{LCHN2} we showed that the $S$ symmetry can account for the lepton asymmetry observed in the universe. The relevant quantity in that calculation was the term
\bea
\left(\tilde{M}_D^{\dagger}  \, \tilde{M}_D\right) _{ij} &=& \left(F_0^\dagger \,V^\dagger_R M_D^{\dagger}\, M_D \,V_R\, F_0\right)_{ij}
\eea
where $\tilde{M}_D$ is the Dirac neutrino mass matrix in the basis where the RH neutrinos are mass eigenstates, $V_R$ is the diagonalizing matrix of $M_R$ and $F_0$ is a phase diagonal matrix so that the eigenvalues of $M_R$ are real positive.

Let's look now at the expression above in the  ``rotated'' basis defined by $W$:
\bea
V^{z\dagger}_R\,  M_D^{z\dagger} M_D^{z}\, V^z_R &=& V^{\dagger}_R\,  M_D^{\dagger}\, M_D \, V_R.
\eea
Moreover, since the diagonal phase matrix $F_0$ depends on the mass spectrum of $M_R$, it remains the same upon going to  ``rotated'' basis defined by $W$. Thus, the relevant discussion for leptogenesis remains the same in both ``non-rotated'' and ``rotated'' $\mu$-$\tau$ symmetries.

\section{Perturbation on $S^z$}
\subsection{Motivation and Preliminaries}
In \cite{LCHN1,LCHN2}, we introduced a perturbation on the $S$-symmetry in order to deal with its experimentally unacceptable zero value for $\t_z$. In the case of
 $S^z$ symmetry, we can arrange to get $\t_z$ equal any value given in advance. However, $S^z$ predicts that $\t_y$ equals exactly $\pi/4$, which is, albeit experimentally allowable,
 not the best fit of $\t_y$. Moreover, the value of $\t_x$ is determined by the entries of $M^z_\n$ and it would be good if we have a freedom in changing slightly the values
  of angles to account for possible new more precise measurements. This pushes us to consider the effects of perturbing the form of $M_\n^z$ imposed by the $S^z$ symmetry. We carry out now
  a complete analytical analysis of $M^z_\n$ perturbations. In the next section we shall present theoretical realizations justifying the possibility of taking the form of perturbations
  we are considering here. As to the numerical analysis, we shall report in a future work the results of scanning the free parameters of the model and determining the regions of parameter space consistent with data. We shall denote the predictions of $S^z$ symmetry (unperturbed) by an $0$ upper index. Thus we have
  \bea
  M^z_\n &=& M^{0z}_\n + \d M^{z}_\n
  \label{Mzp}
  \eea
 and the diagonalizing matrix $U^z$ is defined such that
 \bea
 \label{Qz}
 U^{z\dagger}\ M^{z*}_\n\, M^z_\n\, U^z  &=&  \mbox{Diag} \left( \left|M_{\n 11}^{z \mbox{\tiny Diag}}\right|^2, \left|M_{\n 22}^{z \mbox{\tiny Diag}}\right|^2, \left|M_{\n 33}^{z \mbox{\tiny Diag}}\right|^2
 \right)
 \eea
 whereas the corresponding matrix for the ``unperturbed'' $S^z$ case Eq. (\ref{Uz}) or Eq. (\ref{Uzgeneral})  is denoted by $U^{0z}$:
 \bea
 \label{U0z}
 U^{0z\dagger}\, M^{0z*}_\n\, M^{0z}_\n\, U^{0z}  &=&  \mbox{Diag} \left( \left|M_{\n 11}^{0z \mbox{\tiny Diag}}\right|^2, \left|M_{\n 22}^{0z \mbox{\tiny Diag}}\right|^2, \left|M_{\n 33}^{0z \mbox{\tiny Diag}}\right|^2  \right)
\eea
We define $I_\eps^z$ as
\bea
\label{UIepsz}
U^z &=& U^{0z}\, \left(1 + I_\eps^z\right).
\eea
Unitarity of $U^z$ and $U^{0z}$ means that $I_\eps^z$ is antihermitian:
\bea
\label{Iepsz}
I_\epsilon^z =
\left(
\begin{array}{ccc}
0 & \epsilon^z _1 & \epsilon^z _2 \\
-\epsilon^{z*}_1 & 0 & \epsilon^z _3 \\
-\epsilon^{z*}_2 & -\epsilon^{z*}_3 & 0
\end{array}
\right).
\eea
Working to first order in the perturbation $\d M^{z}_\n$, we get the condition:
\bea
\label{condIepsz}
i,j \in \left\{1,2,3\right\}, i\neq j,
\left[I^z_\epsilon\, ,\,  M_\nu^{0z\mbox{\tiny Diag}*}\,M_\nu^{0z\mbox{\tiny Diag}}\right]_{ij}=
\left[U^{0z\dagger}\, \left(M_\n^{0z*}\,\d M^z_\n + \d M_\n^{z*}\,M_\n^{0z}\right)  U^{0z}\right]_{ij}. &&
\eea
We shall restrict our perturbations to those originating, within seesaw mechanism, from the following perturbation on $M_D^z$ (we assume that $M_R^z$ is form invariant under $S^z$):
\bea
\label{pertMD}
M_D^z = M_D^{0z} + \d M_D^{z}, &&
\d M_D^{z} = \alpha B_D^z
\left(
\begin{array}{ccc}
0 & 1  & 0\\
0 & 0 & 0 \\
0 & 0 & 0
\end{array}
\right).
\eea
Therefore, up to first order perturbation and after having denoted the generic non vanishing entries of $ \d M^z_\n$ by $\a^z_{ij}$ with suitable indexing,  we get $\d M^z_\n$ as
\bea
\label{pertMn}
 \d M^z_\n = \d M^{z\mbox{\tiny T}}_D\ \left(M_R^z\right)^{-1}\, M_D^{0z \mbox{\tiny T}} + M_D^{0z \mbox{\tiny T}}\,  \left(M_R^z\right)^{-1}\, \d M^{z \mbox{\tiny T}}_D
 & = &
\left(
\begin{array}{ccc}
\a_{11}^z & \a_{12}^z  & \a_{13}^z\\
\a_{12}^z & 0 & 0 \\
\a_{13}^z & 0 & 0
\end{array}
\right).
\eea
We note here that the purpose of adding the perturbation $\d M_\n^z $ is to break the symmetry $S^z$, otherwise no new role can play $M^z_\n$ that $M^{0z}_\n$ can not. This non invariance can be easily verified by checking that we have
\bea
\label{breakS}
S^{z \mbox{\tiny T}}\, \left(\d M_\n^{z}\right)\, S^z &\neq& \d M_\n^{z}.
\eea

\subsection{The  Perturbation Form and the Perturbed Mass Spectrum}
We defined in Eq. (\ref{Uz}) the relation between the ``rotated'' unperturbed and the ``non-rotated'' unperturbed bases for the diagonalizing matrices:
\bea
\label{U0z}
U^{0z} &=& W U^0.
\eea
The expressions of $\left(U^0,\, U^{0z}\right)$ are given respectively in Eqs. (\ref{U},\ref{Uzgeneral}).
We note that if  the perturbations in the rotated and non-rotated cases are related by the same similarity transformation defined by $W$ Eq. (\ref{W}), then $I^z_\eps$ is invariant with respect to change of basis, i.e. it is $z$-independent and equal to  $I_\eps$  corresponding to perturbing the non-rotated diagonalizing matrix $U^0$. In fact, defining $U = W^{\mbox{\tiny T}}\, U^z$,
and writing it in the form $U = U^0\, \left(1+ I_\eps\right)$, we see from
\bea
\d M^z_\n = W\, \d M_\n\, W^{\mbox{\tiny T}},&& M^{0z}_\n = W \, M^0_\n\, W^{\mbox{\tiny T}},\;\; U^{0z} = W\, U^0,   \nonumber \\
&\Rightarrow & \nn  \\
 M_\nu^{0z\mbox{\tiny Diag}*}\,M_\nu^{0z\mbox{\tiny Diag}} &=& M_\nu^{0\mbox{\tiny Diag}*}\,M_\nu^{0\mbox{\tiny Diag}},\nn\\
 U^{0z\dagger}\, \left(M_\n^{0z*}\,\d M^z_\n + \d M_\n^{z*}\,M_\n^{0z}\right)\,  U^{0z} & =&  U^{0\dagger}\, \left(M_\n^{0*}\,\d M_\n + \d M_\n^{*}\,M_\n\right)\,  U^{0},
\label{cond}
\eea
and from Eq. (\ref{condIepsz}),  that $I_\eps$ satisfies the same characterizing equation as  $I^z_\eps$.
We thus deduce a method to compute $I_\eps^z$, which is necessary to evaluate the effects of the perturbation $\d M^z_\n$ in the ``rotated' basis. This method consists in writing down the corresponding
perturbation in the ``non-rotated'' basis in the form $\d M_\n = W^T \d M^z_\n W$, and then solving in this latter basis the defining equation:
\bea
\label{condIepsz1}
i,j \in \{1,2,3\},\, i\neq j,\,
\left[I_\epsilon\, ,\,  M_\nu^{0\mbox{\tiny Diag}*}\,M_\nu^{0\mbox{\tiny Diag}}\right]_{ij}=
\left[U^{0\dagger}\, \left(M_\n^{0*}\, W^T \,\d M^z_\n\, W + W^T\, \d M^{z*}_\n\, W \,M_\n^{0}\right)\,  U^{0}\right]_{ij}. &&
\eea
Then the resulting $I_\eps$ is also the correct answer in our rotated basis. We note in addition that the breaking of the symmetry by the perturbation Eq. (\ref{breakS}) is valid in both rotated and
non-rotated bases.

We shall work out now an easier equivalent method to find $I_\eps^z$, which originates from the generic perturbation $\d M^z_\n$, by diagonalizing $\left(M_\n^z\right)$ rather than $\left(M_\n^{z*}\, M_\n^z\right)$.
We have
\bea
 \label{Uzm}
 U^{z\mbox{\tiny T}}\, M^z_\n\, U^z  &=&  \mbox{Diag} \left( M_{\n\, 11}^{z \mbox{\tiny Diag}}, M_{\n\, 22}^{z \mbox{\tiny Diag}}, M_{\n\, 33}^{z \mbox{\tiny Diag}} \right),
 \eea
 and
 \bea
 \label{U0z1}
 U^{0z\mbox{\tiny T}}\, M^{0z}_\n\, U^{0z}  &=&  \mbox{Diag} \left( M_{\n 11}^{0z \mbox{\tiny Diag}},\, M_{\n 22}^{0z \mbox{\tiny Diag}},\, M_{\n 33}^{0z \mbox{\tiny Diag}}  \right).
\eea
Working now to first order in perturbation, we get
\bea
\label{qmq=}
U^{z\mbox{\tiny T}}\, M^z_\n\, U^z &=& M_\nu^{0z\mbox{\tiny Diag}} -I^{z*}_\eps\, \,M_\nu^{0z\mbox{\tiny Diag}}  +  M_\nu^{0z\mbox{\tiny Diag}} \, I^{z}_\eps + U^{0z\mbox{\tiny T}}\, \d M^z_\n\, U^{0z}.
\eea
Expressing the fact that the LHS of Eq. (\ref{qmq=}) and its complex conjugate are diagonal, we obtain  the  following equations which allow us to determine $I_\eps^z$:\footnote{The second line is just the complex conjugate of the first line. One could limit oneself to the first line which represents $3$ complex equations in $3$ complex unknowns $\eps^z_i, i=1,2,3$. However,
it is simpler to treat $\eps^{z*}_i$ as independent from $\eps^z_i$, and get a linear system of $6$ equations in $6$ unknowns. Eventually, we checked that
the obtained solutins are consistent in that $\eps^{z*}_i$ is indeed the complex cnjugate of $\eps^{z}_i$.}
\be
\label{condIepsz2}
\begin{array}{lll}
i,j \in \{1,2,3\},\, i\neq j,\, &&
\left\{
\begin{array}{lll}
\left[-I^{z*}_\eps\, \,M_\nu^{0z\mbox{\tiny Diag}}  +  M_\nu^{0z\mbox{\tiny Diag}}\, I^{z}_\eps + U^{0z\mbox{\tiny T}}\, \d M^z_\n\, U^{0z} \right]_{ij} = 0, &&\nn\\\\
\left[-I^{z}_\eps\, \,M_\nu^{0z*\mbox{\tiny Diag}}  +  M_\nu^{0z*\mbox{\tiny Diag}} I^{z*}_\eps + U^{0z*\mbox{\tiny T}} \d M^{z*}_\n U^{0z*} \right]_{ij} = 0.&&
\end{array}
\right.
\end{array}
\ee
The resulting linear system of six equations in the unknown $\left(\eps_1^z,\,\eps_2^z,\,\eps_3^z,\,\eps^{z*}_1,\,\eps^{z*}_2,\,\eps^{z*}_3\right)$ can be solved and the solutions are reported in Appendix (B) Eq. (\ref{epsilon}). Substituting  these  solutions into Eq. (\ref{qmq=}), we get, for the perturbation $\d M_\n^{z}$ given in Eq. (\ref{pertMn}), the following (recalling $M_\n^{0z\,\mbox{\tiny Diag}}= M_\n^{0\,\mbox{\tiny Diag}}$):
\bea \label{Mspeczp}
M^{z\,\mbox{\tiny Diag}}_{\n\;11} &=& M^{0\,\mbox{\tiny Diag}}_{\n\;11} + \alpha^z_{11}\, c_z^2\, c_{\varphi}^2 - \sqrt{2}\, \left(\alpha_{12}^z - \alpha_{13}^z\right)\,  c_{\varphi} \,s_{\varphi}\, c_z\,
e^{-i\d} -\sqrt{2}\, \left(\alpha_{12}^z + \alpha_{13}^z\right)\, c_{\varphi}^2\, s_z\, c_z, \nonumber \\
M^{z\,\mbox{\tiny Diag}}_{\n\;22} &=& M^{0\,\mbox{\tiny Diag}}_{\n\;22} + \alpha^z_{11}\, c_z^2\, s_{\varphi}^2\, + \sqrt{2}\, \left(\alpha_{12}^z - \alpha_{13}^z\right)\,  c_{\varphi}\, s_{\varphi}\, c_z\,
e^{-i\d} -\sqrt{2}\,  \left(\alpha_{12}^z + \alpha_{13}^z\right)\, s_{\varphi}^2\, s_z\, c_z, \nonumber\\
M^{z\,\mbox{\tiny Diag}}_{\n\;33} &=& M^{0\,\mbox{\tiny Diag}}_{\n\;33} + \alpha^z_{11}\, s_z^2 + \sqrt{2}\, \left(\alpha_{12}^z + \alpha_{13}^z\right)\,  s_z\, c_z.
\eea
The two procedures for determining $I_\eps^z$ via Eq. (\ref{condIepsz}) or Eq. (\ref{condIepsz2}) are shown, in Appendix (B), to be equivalent, as expected.

\subsection{Determining the Resulting Mixing and Phase Angles after Perturbation}
Having now determined the matrix $I_\eps^z$ (Eq. (\ref{epsilon})),  then the resulting mixing matrix, denoted by $U_{\eps}$, would be
\be
U_\eps = U^{0z}\, \left(1 + I_\eps^z\right)\,\mbox{Diag}\left(e^{-i\phi_1},e^{-i\phi_2},e^{-i\phi_3}\right),
\label{ueps}
\ee
where $\phi_i = \displaystyle{{1\over 2}}\, \mbox{Arg}\left(M_{\n\, ii}^{z\,\mbox{\tiny Diag}}\right)$. The multiplication by the determined diagonal phase matrix is required in order to make the eigenvalues of $M_\n$ real and positive. Before extracting the resulting mixing and phase angles, one needs to carry out a further rephasing in order to make the third column of $U_\eps$ real, so that to be consistent with the adopted parametrization of Eq. (\ref{defv}). Thus, we rephase the fields of the charged leptons as,
\bea
e\,\rightarrow e^{i\psi_1}\,e,\;\; \mu\,\rightarrow\, e^{i\psi_2}\, \mu, \;\; \tau\,\rightarrow\, e^{i\psi_3}\, \tau,\;\;\mbox{where}\;\; \psi_i = \mbox{Arg}\left[U_\eps\left(i,3\right)\right],
\label{lepph}
\eea
The full expressions for all entries of the matrix $U_\eps$ are presented in Eq. (\ref{uepsent}) in Appendix (B).

 Identifying now $U_{\eps}$, after having suitably rephased the charge leptonts, with $V_{\mbox{\tiny PMNS}}^*$, we can extract the angles. We list now the first order approximations for $\t_x$, $\t_y$ and $\t_z$  whereas the full expressions are listed in the Appendix (B) Eq. (\ref{mix_xyz}),
\bea
\label{1orxyz}
t_x &\simeq& t_\varphi\, \left|1+ \displaystyle{{1\over t_\varphi}}\, \eps_1^{z} + t_\varphi\, \eps_1^{z*} - \displaystyle{{t_{z0}\over s_\varphi}}\, \eps_3^{z*} + \displaystyle{{t_{z0}\over c_\varphi}}\, \eps_2^{z*}\right|,\nn\\
t_y &\simeq& \left|1-2\, \eps_2^z\, \displaystyle{{s_\varphi\over c_{z0}}}\, e^{-i\xi} + 2\, \eps_3^z \displaystyle{{c_\varphi\over c_{z0}}}\, e^{-i\xi} \right|,\nn\\
s_z &\simeq& s_{z0}\,\left|1 + \displaystyle{{c_\varphi \over t_{z0}}}\, \eps_2^{z}  + \displaystyle{{s_\varphi \over t_{z0}}}\, \eps_3^{z} \right|,\nn\\
\eea
where $z0$ is the angle which determines the rotated symmetry before perturbation, whereas $z$ corresponds to the perturbed texture.

As to the Majorana phase angles, which by convention belong to the first and second quadrants, they are determined to be:
\bea
\label{rhosig}
\rho &=& \pi - \mbox{Arg}\left[\displaystyle{{c_{z0}\, c_\varphi - c_{z0}\, s_\varphi\, \eps_1^{z*} -s_{z0}\, \eps_2^{z*}\over s_{z0}+c_{z0}\, c_\varphi\, \eps_2^{z}+c_{z0}\, s_\varphi \eps_3^{z}}}\right]
-\displaystyle{{1\over 2}}\,\mbox{Arg}\left(M_{\n\, 33}^{z\,\mbox{\tiny Diag}}\,M_{\n\, 11}^{z\,\mbox{\tiny Diag}*}\right),\nn\\
\sigma &=& \pi - \mbox{Arg}\left[\displaystyle{{c_{z0}\, s_\varphi + c_{z0}\, c_\varphi\, \eps_1^z -s_{z0}\, \eps_3^{z*}\over s_{z0}+c_{z0}\, c_\varphi\, \eps_2^{z}+c_{z0}\, s_\varphi\, \eps_3^{z}}}\right]
-\displaystyle{{1\over 2}}\,\mbox{Arg}\left(M_{\n\, 33}^{z\,\mbox{\tiny Diag}}\,M_{\n \,22}^{z\,\mbox{\tiny Diag}*}\right).
\eea
Finally, and after determining the mixing and Majorana phases, we can get the Dirac phase $\d$ by solving an equation which results upon equating an entry of $V^*_{\mbox{\tiny PMNS}}$ involving $\d$  with the corresponding one of $U_\eps$  (look for example at Eq. (\ref{delta}) in Appendix B).

\section{Realization of perturbed rotated textures}

It is important to find theoretical realizations for the perturbed textures, assuming at the level of the Lagrangian exact symmetries, some of which are broken spontaneously. We need
to parameterize the perturbations on $M^{0z}_\n$, which we shall assume originating from perturbations on only $M^{0z}_D$, so we need to parameterize the latter perturbations also.
As to $M_R^z$ we shall assume that it is invariant under $S^z$. We find two parameters, $\chi$ and $\xi$ for the perturbations in $M^{z}_\n$, that we shall dis-entangle in our future numerical work scanning
the parameter space and contrasting to data, as was done in the case of $S$ symmetry \cite{LCHN2}. Thus the question arises whether or not we can find a theoretical realization for the perturbed texture where one of the parameters, say $\chi$ only is present. In \cite{LCHN2}, we carried out this task for $S$-symmetry and we aim now to generalize this to $S^z$-symmetry, which, as we shall see, is not a trivial task.

\subsection{Parameterizing the Perturbations}
The general form for a symmetric matrix invariant under $S^z$ is given in Eqs. (\ref{Mnz}, \ref{Mzncons1}). We rewrite them here for the unperturbed matrix $M_\n^{0z}$:
\bea
M_{\n\,13}^{0z} &=& -\left(1-2\,t_{z0}^2\right) M_{\n\,12}^{z0} -\sqrt{2}\, t_{z0}\, \left(M_{\n\,11}^{0z}-M_{\n\,22}^{0z}-M_{\n\,23}^{0z}\right),\nn  \\
M_{\n\,33}^{0z} &=& \left(1-2\,t_{z0}^2\right) M_{\n\,22}^{0z} + 2\, t_{z0}^2\, \left(M_{\n\,11}^{0z} - M_{\n\,23}^{0z}\right) + 2\, \sqrt{2}\, t_{z0}\, \left(1-t_{z0}^2\right)\, M_{\n\,12}^{0z}.
\label{Mzncons2}
\eea

 As there are two constraints, Eq. (\ref{Mzncons2}), on the symmetric matrix obeying $S^z$ symmetry, we thus define two parameters which determine the perturbation by measuring the deviations from these two constraints:
\bea
\label{per_chixi}
\chi &=& \frac{M_{\n\, 13}^z-\left[-\left(1-2\,t_z^2\right)\, M_{\n\, 12}^z -\sqrt{2}\, t_z\, \left(M_{\n\, 11}^z-M_{\n\, 22}^z-M_{\n\, 23}^z\right)\right]}{M_{\n\, 12}^z},\nn\\
\xi &=& \frac{M_{\n\, 33}^z-\left[\left(1-2\,t_z^2\right)\, M_{\n\, 22}^z + 2\, t_z^2\, \left(M_{\n\, 11}^{z}-M_{\n\, 23}^z\right) + 2\, \sqrt{2}\, t_z\, \left(1- t_z^2\right) \,M_{\n\, 12}^z\right]}{M_{\n\, 33}^z},
\eea

The invariance of $M^{0z}_D$ under $S^z$ forces it to be parameterized in a such a way as
presented through Eqs. (\ref{MzD}--\ref{Mzdcon1}). We rewrite it here in the following form:
\bea
M_{D\,13}^{0z} &=& - M_{D\,12}^{0z} + \sqrt{2}\,t_{z0}\,\left(M_{D\,23}^{0z} + M_{D\,22}^{0z} - M_{D\,11}^{0z}\right) + 2\,t_{z0}^2\, M_{D\,21}^{0z},\nn\\
M_{D\,32}^{0z} &=& M_{D\,23}^{0z} + \sqrt{2}\,t_{z0}\,\left(M_{D\,21}^{0z}  - M_{D\,12}^{0z}\right),\nn\\
M_{D\,33}^{0z} &=& \left(1-2\,t_{z0}^2\right)\,M_{D\,22}^{0z} - 2\,t_{z0}^2\,\left(M_{D\,23}^{0z}  - M_{D\,11}^{0z}\right) + \sqrt{2}\,t_{z0}\,M_{D\,12}^{0z} +\sqrt{2}\,t_{z0}\,\left(1- 2\,t_{z0}^2\right)\, M_{D\,21}^{0z},\nn\\
M_{D\,31}^{0z} &=& -\left(1- 2\,t_{z0}^2\right)\, M_{D\,21}^{0z} + \sqrt{2}\,t_{z0}\,\left(M_{D\,23}^{0z} + M_{D\,22}^{0z} - M_{D\,11}^{0z}\right).
\label{Mzdcon2}
\eea
Moreover, in Appendix (A), we restate this form of a general matrix invariant under $S^z$ (called there $M_D^{z}$) in Eq. (\ref{Mzdcon3}), while the relations between the rotated and the non-rotated parameters are written in Eqs. (\ref{relADzAD}--\ref{relADADz}). Having four constraints on $M^{0z}_D$ implied  by $S^z$ invariance, as presented in Eq. (\ref{Mzdcon2}), leads naturally to require four parameters in order to quantify the deviation from $S^z$ symmetry. However, for simplicity and flexibility purposes, one can work by taking the mass matrix $M^z_D$ as bearing only two perturbation parameters $\a$ and $\b$ in the form:
\bea
\label{MzpD}
M^{z}_D &=& \,
\left(\begin{array}{ccc}
A^z_D & B^z_D (1+\a) & B'^z_D\\
E^z_D (1+\b) & C^z_D & D^z_D \\
E'^z_D & D'^z_D & C'^z_D
\end{array}
\right),
\eea
where the set of parameters $\left\{ B'^z_D,\, C'^z_D,\,D'^z_D,\, E'^z_D\right\}$ are related to the set $\left\{A^z_D,\, B^z_D,\, C^z_D,\, D^z_D,\, E^z_D\right\}$ through the
same relations given in Eq. (\ref{Mzdcon1}). This fixes the prameters $\a, \b$ for a given perturbed mass matrix $M_D^z$.

The ``perturbed'' $M^z_D$, as given by Eq. (\ref{MzpD}), satisfies the following relations, which clarify how the parameters $\a, \b$ measure the deviations from the constraints of Eq. (\ref{Mzdcon2}):
\bea
\label{relMzpD}
&&M^z_{D\,13} + M^z_{D\,12} + \sqrt{2}\, t_{z0}\, \left( M^z_{D\,11}- M^z_{D\,22} -M^z_{D\,23}\right) - 2\, t_{z0}^2\, M^z_{D\,21} =
\a\, B^z_D -2\, \b\, t_{z0}^2\, E^z_D,\nn \\
&&M^z_{D\,33} - \left(1-2\,t_{z0}^2\right)\, M^z_{D\,22} +2\, t_{z0}^2\, \left( M^z_{D\,23}- M^z_{D\,11}\right) - \sqrt{2}\, t_{z0}\, M^z_{D\,12} - \sqrt{2}\,t_{z0}\,\left(1-2\,t_{z0}^2\right)\,M^z_{D\,21}  =\nn\\
&&-\sqrt{2}\,t_{z0}\,\left[\a\, B^z_D + \b\,\left(1- 2\, t_{z0}^2\right)\, E^z_D\right],\nn\\
&&M^z_{D\,32} - M^z_{D\,23} - \sqrt{2}\, t_{z0}\, \left( M^z_{D\,21}- M^z_{D\,12}\right) = \sqrt{2}\, t_{z0}\, \left(\a\, B^z_D - \b\, E^z_D\right),\nn \\
&&M^z_{D\,31} + M^z_{D\,21} + \sqrt{2}\, t_{z0}\, \left( M^z_{D\,11}- M^z_{D\,22} -M^z_{D\,23}\right) - 2\, t_{z0}^2\, M^z_{D\,21} =
\b\, E^z_D\, \left(1-2\,t_{z0}^2\right).
\eea
The perturbations induced by $\a$ and $\b$ in $M_D^z$ would be transmuted into $M_\n^z$ through the seesaw mechanism described in Eq. (\ref{seesaw}), and one
 can compute the corresponding perturbations parameters $\chi$ and $\xi$ which, to first order in $\a$, $\b$ and $s_z$, turn out to be
{\small
\bea
&& \chi = \frac{-\a\, B^z_D\, \left(C^z_D - D^z_D\right) \left[ A^z_R \left(C^z_R - D^z_R\right) - 2\,B^{z2}_R\right] -\b\, E^z_D\, \left(C^z_R+D^z_R\right)
\left[ A^z_D\, \left(C^z_R - D^z_R\right) - 2\,B^z_R\, B^z_D\right]  }
{\left(C^z_R + D^z_R\right)\, \left[ B^z_D\, A^z_R\, \left(D^z_D - C^z_D\right)  +
E^z_D\, A^z_D\, \left(D^z_R - C^z_R\right) + B^z_R\, \left(2\, B^z_D\, E^z_D + A^z_D\, C^z_D - A^z_D\, D^z_D\right)\right]},\nn\\\nn\\
&& \xi =\frac{-2\, \b\, E^z_D\, \left(C^z_R + D^z_R\right)\, \left[ E^z_D\, \left(C^z_R - D^z_R\right) + B^z_R\,\left(D^z_D - C^z_D\right)\right]  }
{\left(A^z_R\, C^z_R - B^{z2}_R\right)\, \left(C^{z2}_D + D^{z2}_D\right)  + \left(C^z_R + D^z_R\right)\, \left[ 2\, B^z_R\, E^z_D\, \left(D^z_D + C^z_D\right)  +  E^{z2}_D\,  \left(C^z_R - D^z_R\right) -2\, C^z_D\, D^z_D\, \left( A^z_R\, D^z_R +  B^{z2}_R\right)\right]}.\nn\\
\label{chixiexp}
\eea
}
We see directly, up to this given order, that when $\b = 0$ then $\xi = 0$. As we seek in this section a realization for a dis-entangled perturbation parameterized solely by $\chi$, then we shall
look for a realization of $M^z_D$ with $\b = 0$.

In \cite{LCHN2} we found a realization of the ``dis-entangled'' perturbation, due only to $\chi$ and not to $\xi$, assuming exact $S$-symmetry but at the expense of extending the symmetry
and adding new matter. Here, we shall do the same but with the symmetry $S^z$. In order to find the $S^z$-transformations knowing the corresponding $S$-ones, we use the following rule of thumb:
\bea
\label{ruleofthumb}
\left(\mbox{rotated symmetry element}\right) &=& W\, \left(\mbox{ non-rotated symmetry element }\right)\, W^{\mbox{\tiny T}}
\eea
As in \cite{LCHN2}, we present two ways to get a perturbed $M^z_\n$ with $\xi=0$, the first one assuming a $S^z  \times  Z_2^2 $ symmetry, whereas the symmetry in the other  way is $S^z \times  Z_8 $.

\subsection{ $S^z \times Z_2 \times Z_2^\prime$-flavor symmetry}

\begin{itemize}
\item{\bf Matter content and symmetry transformations}

We have three SM-like Higgs doublets ($\phi_i$, $i=1,2,3$) giving mass to the charged leptons and another three Higgs doublets ($\phi^\prime_i$, $i=1,2,3$) for the
Dirac neutrino mass matrix.  All the fields remain unchanged under $Z_2^\prime$ except the fields $\phi^\prime$ and $\n_R$ which are multiplied by $-1$, so that we
assure that neither $\phi$ can contribute to $M_D$, nor $\phi^\prime$ to $M_l$.  We had in \cite{LCHN2} the assignment of the fields under the $S$-symmetry, and so by the rule of thumb we get the following transformations.\\
The transformations under $Z_2$ are
\bea
\n_R &\stackrel{Z_2}{\longrightarrow}& W\; \mbox{Diag}\left(1,-1,-1\right)\; W^T \n_R ,\;\; \phi^\prime \stackrel{Z_2}{\longrightarrow} W\; \mbox{Diag}\left(1,-1,-1\right) W^T\; \phi^\prime,\nn\\
L & \stackrel{Z_2}{\longrightarrow}& W\; \mbox{Diag}\;\left(1,-1,-1\right)\; W^T \; L,\;\; l^c \stackrel{Z_2}{\longrightarrow} W\; \mbox{Diag}\left(1,1,-1\right)\; W^T\;  l^c ,\nn\\
 \phi &\stackrel{Z_2}{\longrightarrow}& W\; \mbox{Diag}\left(1,-1,-1\right)\; W^T\; \phi.
 \label{uno_z2}
 \eea
The transformation under $S^z$ are
\bea
\n_R &\stackrel{S^z}{\longrightarrow}& S_{\n_R}^z\; \n_R = S^z\; \n_R ,\;\;\;\; \phi^\prime \stackrel{S^z}{\longrightarrow} S^z_{\phi^\prime}\; \phi^\prime = W\; \mbox{Diag}\left(1,1,-1\right)\; W^T\; \phi^\prime,\nn\\
L &\stackrel{S^z}{\longrightarrow}& S^z_L\; L = S^z\; L, \hspace{1cm} l^c \stackrel{S^z}{\longrightarrow} W\; \mbox{Diag}\left(1,1,1\right)\; W^T\; l^c =  l^c,\nn\\ \phi &\stackrel{S}{\longrightarrow}& S^z_\phi\; \phi = S^z \; \phi.
\label{uno_sz}
 \eea
\item{\bf Charged lepton mass matrix-flavor basis}

The Lagrangian responsible for $M^z_l$ is given by:
\bea
 \label{L2}
 {\cal{L}}_2 &=& f^{zj}_{\,ik}\; \overline{L}_i\; \phi_k \; l^c_j \,
 \eea
 The invariance of the Lagrangian under $S^z$ implies the following for the Yukawa couplings $f^{zj}_{\,ik}$:
  \bea
  \label{fzjik}
  S^{z\mbox{\tiny T}}_{L\,li}\; f^{zj}_{\,ik}\; S^z_{\phi\, km}\; S^z_{{l^c}\! jn} &=& f^{zn}_{\,lm}.
  \eea
In order to find $f^{zj}_{\,ik}$, one can start from the known solutions in the case of $S$-symmetry:
\bea
  \label{fjik}
  S^{\mbox{\tiny T}}_{L\,li}\; f^{j}_{\,ik}\; S_{\phi\, km}\; S_{{l^c}\! jn} &=& f^{n}_{\,lm},
  \eea
and expressing the $S$'s in terms of the $S^z$'s in that $S= W^T\, S^z\, W$, we get
 \bea
 \label{step1fz}
 W^{\mbox{\tiny T}}_{l \a}\; S^{z\mbox{\tiny T}}_{L\a \b}\; W_{\b i}\; f^j_{ik}\;
 W^{\mbox{\tiny T}}_{k \gamma}\;  S^z_{\phi \gamma \rho}\; W_{\rho m}\;
 W^{\mbox{\tiny T}}_{j \sigma}\;
 S^z_{{l^c} \sigma \t}\;  W_{\t n} &=& f^n_{lm}.
 \eea
We find that a solution of Eq. (\ref{fzjik}) is given by:
 \bea
 \label{step2fz}
 f^{z\s}_{\b \gamma} &=& W_{\b i}\; f^j_{ik}\; W^{\mbox{\tiny T}}_{k \gamma}\; W^{\mbox{\tiny T}}_{j \sigma}.
 \eea
 Defining the matrices ${\bf f}^{zj}$ and ${\bf f}^{j}$ as the matrices whose $(i,k)$-th entries are respectively $f^{zj}_{ik}$ and $f^j_{ik}$ then we can express Eq. (\ref{step2fz}) as
 \bea
 \label{matfexp}
 {\bf f}^{z\s} &=& W \; {\bf f}^{j} \; W^{\mbox{\tiny T}}\; (W^{\mbox{\tiny T}})_{j \sigma},
 \eea
 which means that the solution for the ``rotated'' basis is obtained by a similarity transformation applied onto the solution for ``non-rotated'' basis, followed by a linear combination
 weighted by $(W^{\mbox{\tiny T}})_{j \sigma}$. Moreover we can re-express the symmetry constraint of Eq. (\ref{fzjik}) in matrix form as a weighted sum of similarity transformations:
 \bea
 \label{fzjikmat}
 S_L^{z\mbox{\tiny T}}\;\; {\bf f}^{z \sigma} \; S^z_\phi \;\; (S^z_{l^c})_{\sigma \Lambda} &=& {\bf f}^{z\Lambda}.
 \eea

 Now, using the results of \cite{LCHN2} where, taking into consideration the invariance under both $S$ and $Z_2$-symmetries, we obtained
\bea
\label{fnon-rotated}
{\bf f}^{1}  =
\left(
\begin {array}{ccc}
A^1 &0& 0\\
0& C^1& D^1\\
0&D^1&C^1
\end {array}
\right),\;
{\bf f}^{2}  =
\left(
\begin {array}{ccc}
A^2 &0& 0\\
0& C^2& D^2\\
0&D^2&C^2
\end {array}
\right),\;
{\bf f}^{3}  =
\left(
\begin {array}{ccc}
0 &B^3& -B^3\\
E^3& 0& 0\\
-E^3&0&0
\end{array}
\right),
\eea
 and applying the similarity transformations by $W$:
 \bea
 \label{similarity}
 {\bf \tilde{f}}^{zj} &=& W \; {\bf f}^j\; W^{\mbox{\tiny T}}
 \eea
 we get the matrices ${\bf \tilde f}^{zj}, j=1,2,3$ whose expressions, in terms of new coefficients
  $\left(A^z_i,\,B^z_i,\, C^z_i,\,D^z_i,\; i=1,2\right)$ and $\left(E^z_3,\, B^z_3\right)$ related to the old coefficients, are given in Appendix C
 Eq. (\ref{tildef}) and Eq. (\ref{tildefecoef}). We follow this by the weighted sum ${\bf f}^{z\sigma} = {\bf \tilde{f}}^{zj}\; W^{\mbox{\tiny T}}_{j\sigma}$ using the expressions of $W$ in Eq. (\ref{W}) to find finally
 \bea
 \label{fmat}
 {\bf f}^{z1} &=& c_z\,{\bf \tilde f}^{z1}  + \displaystyle{{s_z\over \sqrt{2}}}\;{\bf \tilde f}^{z2}  +\displaystyle{{s_z\over \sqrt{2}}}\;{\bf \tilde f}^{z3},\nn \\ \nn\\
 {\bf f}^{z2} &=& -\displaystyle{{s_z\over \sqrt{2}}}\;{\bf \tilde f}^{z1}  + c^2_{{z/ 2}}\,{\bf \tilde f}^{z2} - s^2_{z/2}\, {\bf \tilde f}^{z3},\nn \\ \nn\\
 {\bf f}^{z3} &=& -\displaystyle{{s_z\over \sqrt{2}}}\;{\bf \tilde f}^{z1}  -s^2_{z/2}\, {\bf \tilde f}^{z2} + c^2_{z/2}\,{\bf \tilde f}^{z3} .
  \eea
When the Higgs fields $\phi$'s acquire vevs, and assuming ($v_3 \gg v_1, v_2$) we get to lowest order in $s_z$:
\bea
M_l^z  = v_3
\left(
\begin {array}{ccc}
0 &0& -B^z_3\\
D^z_1& D^z_2 & 0\\
C^z_1&C^z_2&0
\end {array}
\right) &\Rightarrow&
M_l^z\;M_l^{z\dagger}
    = v_3^2
    \pmatrix {|{\bf B}^z|^2  & 0 & 0 \cr
     0 & |{\bf D}^z|^2 & {\bf D}^z \cdot {\bf C}^z\cr
    0 & {\bf C}^z \cdot {\bf D}^z & |{\bf C}^z|^2},
\label{chMLszz2}
\eea
where ${\bf B}^z=\left(0,0,-B^z_3\right)^T$, ${\bf D}^z=\left(D^z_1,D^z_2,0\right)^T$ and ${\bf C}^z=\left(C^z_1,C^z_2,0\right)^T$, and where the dot product is defined as ${\bf D}^z \cdot {\bf C}^z = \sum_{i=1}^{i=3}
D^z_iC^{z*}_{i}$. Under the reasonable assumption that the magnitudes of the Yukawa couplings come in ratios proportional to the lepton mass ratios as $\left|B^z\right| : \left|C^z\right| : \left|D^z\right|
\sim m_e : m_\mu : \m_\tau$, we can show, as was done in \cite{LCHN1}, that this form can be  diagonalized by infinitesimal rotations applied onto the LH charged lepton fields, which justifies working in the flavor basis to a good approximation.

\item{\bf Majorana neutrino mass matrix}\\
The mass term is directly present in the Lagrangian
\bea
 {\cal{L}}_R &=& {1\over 2}\, \n_{Ri}^T\, C^{-1}\, \left(M^z_R\right)_{ij}\, \n_{Rj}. \label{csawLR}
 \eea
The invariance under $Z_2'$ is trivially satisfied while the one under
$S^z\times Z_2$ is more involved. In \cite{LCHN2}, we found a form similar to, say, ${\bf f}^{1}$ in Eq. (\ref{fnon-rotated}), which was invariant under $S\times Z_2$, and so the corresponding
form in the ``rotated'' basis for $M_R^z$ would be similar to ${\bf \tilde{f}}^1 = W\; {\bf f}^{1}\; W^T$, i.e.
 that $M_R^z$ would
assume the following form,
\bea
\label{mrs}
M^z_R & = &
\left(
\begin {array}{ccc}
A^z_R &B^z_R& B^z_R\\
B^z_R& C^z_R& D^z_R\\
B^z_R&D^z_R &C^z_R
\end {array}
\right),\;\; \mbox{where}\;\; B^z_R = -\displaystyle{{t_{2z}\over 2\,\sqrt{2}}}\, \left(A^z_R - C^z_R -D^z_R\right).
\eea
One can check that $M^z_R$ above does satisfy the constraints of Eq. (\ref{Mnz}, \ref{Mzncons1}) showing that $M^z_R$ is $S^z$-invariant.


\item{\bf Dirac neutrino mass matrix}

The Lagrangian responsible for the neutrino mass matrix is
\bea
 \label{L_D}
 {\cal{L}}_D &=& g^{zk}_{ij}\; \overline{L}_i\; \tilde{\phi^\prime}_k\;  \n_{Rj}, \;\; \mbox{where}\;\; \tilde{\phi^\prime} = i\, \s_2\,  \phi^{\prime *}.
 \eea
In a similar manner to our discussion in the above item about the charged lepton mass matrix, we find that invariance under $S^z$ implies the following constraint
on Yukawa couplings (c.f. Eq. (\ref{fzjik})):
\bea
  \label{gzjik}
  S^{z\mbox{\tiny T}}_{L\,li}\; g^{zk}_{ij}\; S^z_{\n_R\, jm}\; S^z_{\tilde{\phi'} kn} &=& g^{zn}_{\,lm},
  \eea
which can be written in an equivalent matrix form similar to Eq. (\ref{fzjikmat}) as
\bea
 \label{gzjikmat}
 S_L^{z\mbox{\tiny T}}\;\; {\bf g}^{z \sigma} \; S^z_{\n_R} \;\; (S^z_{\tilde{\phi'}})_{\sigma \Lambda} &=& {\bf g}^{z\Lambda}.
 \eea where the matrix ${\bf g}^{zj}$ has $g^{zj}_{ik}$  at its $(i,k)$-th entry.

Again, knowing the solution ${\bf g}$ for ``non-rotated'' case we can get the corresponding one for the rotated case as,
\bea
 \label{matgexp}
 {\bf g}^{z\s} &=& W \; {\bf g}^{j} \; W^{\mbox{\tiny T}}\; (W^{\mbox{\tiny T}})_{j \sigma},
 \eea

In \cite{LCHN2}, we found the expressions of $g^j_{ik}$ taking into consideration the $S$ and $Z_2$ symmetries:
\bea
\label{gnon-rotated}
{\bf g}^{1}  =
\left(
\begin {array}{ccc}
{\cal A}^1 &0& 0\\
0& {\cal C}^1& {\cal D}^1\\
0&{\cal D}^1&{\cal C}^1
\end {array}
\right),\;
{\bf g}^{2}  =
\left(
\begin {array}{ccc}
0 &{\cal B}^2& -{\cal B}^2\\
{\cal E}^2& 0& 0\\
-{\cal E}^2&0&0
\end {array}
\right),\;
 {\bf g}^{3}  =
 \left(
\begin {array}{ccc}
0 &{\cal B}^3& {\cal B}^3\\
{\cal E}^3& 0& 0\\
{\cal E}^3&0&0
\end {array}
\right).
\eea
We apply now the similarity transformations by $W$:
 \bea
 \label{similarityg}
 {\bf \tilde{g}}^{zk} &=& W \; {\bf g}^k \; W^{\mbox{\tiny T}},
 \eea
 and we get the matrices ${\bf \tilde g}^{zk}, k=1,2,3$ whose expressions, in terms of new coefficients
  (${\cal A}^z_i,{\cal B}^z_i,{\cal C}^z_i,{\cal D}^z_i, {\cal E}^z_i $, $i=1,2,3$) related to the old coefficients, are given in Appendix C
 (Eq. (\ref{tildeg}) and Eq.\ref{tildegecoef})). We follow this by the weighted sum ${\bf g}^{z\sigma} = {\bf \tilde{g}}^{zk}\; W^{\mbox{\tiny T}}_{k\sigma}$ (c.f. Eq. (\ref{fmat}) replacing
  ${\bf f}^{z}$ by ${\bf g}^{z}$).

Upon acquiring vevs ($v_i^\prime$, $i=1,2,3$) for the Higgs fields ($\phi^\prime_i$), we get, up to leading order in $s_z$, for Dirac neutrino mass matrix the form:
\bea
M_D = {\bf g}^{z \sigma}  v_\sigma^\prime &=&\left(
\begin {array}{ccc}
v_1'\, {\cal A}^z_1 & v_2'\, {\cal B}^z_2 + v_3'\, {\cal B}^z_3 & -v_2'\, {\cal B}^z_2 + v_3'\, {\cal B}^z_3\\
v_2'\, {\cal E}^z_2 + v_3'\, {\cal E}^z_3 & v_1'\, {\cal C}^z_1 & v_1'\, {\cal D}^z_1 \\
-v_2'\, {\cal E}^z_2 + v_3'\, {\cal E}^z_3 & v_1'\, {\cal D}^z_1 & v_1'\, {\cal C}^z_1
\end{array}
\right),
\label{MD1}
\eea
which can be matched, up to leading order of $s_z$, with the form of Eq. (\ref{relMzpD}) to yield,
\bea
\a = \frac{2v^\prime_3 {\cal B}^z_3}{v^\prime_2 {\cal B}^z_2 - v^\prime_3 {\cal B}^z_3} &,& \b = \frac{2v^\prime_3 {\cal E}^z_3}{v^\prime_2 {\cal E}^z_2 - v^\prime_3 {\cal E}^z_3}.
\label{alfabeta}
\eea
If the vevs satisfy $v^\prime_3 \ll v^\prime_2$ and the Yukawa couplings are of the same order, then we get perturbative  parameters $\a, \beta \ll 1$.
These perturbative parameters resurface as perturbative parameters for $M^z_\n$ Eq. (\ref{per_chixi}). Although we do not get in general
 disentanglement of the perturbations ($\xi =0$), however, for specific choices of Yukawa couplings, for e.g. ${\cal E}^z_3=0$ leading to $\b=0$ and hence $\xi =0$, we get this disentanglement, where only $\chi$ is not equal to zero and is given by Eq. (\ref{chixiexp}) with $B^z_R=0$ to lowest order.
\end{itemize}

\subsection{ $S^z \times Z_8 $-flavor symmetry}
Here, and as was the case in \cite{LCHN2}, we shall find a realization that gives $\b=0$ regardless of the Yukawa couplings values.
\begin{itemize}
\item{\bf Matter content and symmetry transformations}

  We have the left doublets ($L_i$, $i=1,2,3$), the RH charged singlets ($l^c_j$, $j=1,2,3$), the RH neutrinos ($\n_{Rj}$, $j=1,2,3$) and the SM-Higgs three doublets ($\phi_i$, $i=1,2,3$) responsible for the charged lepton masses. We have also four Higgs doublets ($\phi^\prime_j$, $j=1,2,3,4$) leading  to Dirac neutrino mass matrix, and two Higgs singlet scalars
($\Delta_k$, $k=1,2$) related to Majorana neutrino mass matrix. We denote the octic root of the unity by $\omega=e^{\frac{i\pi}{4}}$. The fields transform according to the rule of thumb Eq. (\ref{ruleofthumb}) as follows.\\
The transformations under $S^z$ are
\bea
L &\stackrel{S^z}{\longrightarrow}& S^z_L\; L = S^z \;L,\;\; l^c \stackrel{S^z}{\longrightarrow} W\; \mbox{Diag}\left(1,1,1\right)\; W^{\mbox{\tiny T}}\; l^c = l^c,\;\; \phi \stackrel{S^z}{\longrightarrow} S^z_\phi\; \phi = S^z\; \phi,\nn\\
\nn\\
 \n_R &\stackrel{S^z}{\longrightarrow}& S^z_{\n_R}\; \n_R = S^z\; \n_R,\;\;
  \phi^\prime \stackrel{S^z}{\longrightarrow} W_{\mbox{\tiny 4ext}}\;\mbox{Diag}\left(1,1,1,-1\right)\; W^\dagger_{\mbox{\tiny 4ext}}\; \phi^\prime,\nn\\ \nn\\
   \Delta  &\stackrel{S^z}{\longrightarrow}& W_{\mbox{\tiny 2ext}}\; \mbox{Diag}\left(1,1\right)\; W^\dagger_{\mbox{\tiny 2ext}}\; \Delta = \Delta,\;\;
   \tilde{\phi^\prime}  \stackrel{S^z}{\longrightarrow} W_{\mbox{\tiny 4ext}}\;\mbox{Diag}\left(1,1,1,-1\right)\; W^\dagger_{\mbox{\tiny 4ext}}\; \tilde{\phi^\prime}.
     \label{sec_sz}
\eea
The transformation under $Z_8$ are
\bea
\label{sec_z8}
L &\stackrel{Z_8}{\longrightarrow}& W\;\mbox{Diag}\left(1,-1,-1\right)\; W^{\mbox{\tiny T}}\;  L,\;\; l^c \stackrel{Z_8}{\longrightarrow}\; W\; \mbox{Diag}\; \left(1,1,-1\right)\; W^{\mbox{\tiny T}}\; l^c,  \nn\\\nn\\
 \phi &\stackrel{Z_8}{\longrightarrow}& W\; \mbox{Diag}\;\left(1,-1,-1\right)\; W^{\mbox{\tiny T}}\; \phi,\;\;
 \n_R \stackrel{Z_8}{\longrightarrow} W \; \mbox{Diag}\left(\omega,\omega^3,\omega^3\right)\; W^{\mbox{\tiny T}}\; \n_R,\nn\\
 \nn\\
 \phi^\prime &\stackrel{Z_8}{\longrightarrow}& W_{\mbox{\tiny 4ext}}\;\mbox{Diag}\;\left(\omega,\omega^3,\omega^7,\omega^3\right)\; W^\dagger_{\mbox{\tiny 4ext}}\; \phi^\prime,\;\;
 \Delta  \stackrel{Z_8}{\longrightarrow} W_{\mbox{\tiny 2ext}}\; \mbox{Diag}\left(\omega^6,\omega^2\right)\; W^\dagger_{\mbox{\tiny 2ext}}\; \Delta,
 \nn\\\nn\\
 \tilde{\phi^\prime}  &\stackrel{Z_8}{\longrightarrow}& W_{\mbox{\tiny 4ext}}\; \mbox{Diag}\left(\omega^7,\omega^5,\omega,\omega^5\right)\; W^\dagger_{\mbox{\tiny 4ext}}\;  \tilde{\phi^\prime}.
 \eea

We note here that we need to extend the symmetry $S^z$ to the case of two and four dimensional representations, in that we need to define the action of the element $W$ of the ``rotations'' group  over the $2$-dim $\Delta$-field and over the $4$-dim $\phi^\prime$-field.
We also note that we use $W^\dagger$ rather than $W^{T}$, since it is the inverse $W^{-1}$ which is involved in the definition of the similarity transformation from the ``non-rotated'' to the ``rotated'' bases. For a unitary complex matrix, it is $W^\dagger$ which represents the inverse and not $W^T$.

The extension of the $W$-action from the fundamental representation of the rotations group acting on $3$-dim space to $4$-dim space is carried in the simplest way by embedding the $3$-dim rotation into a $4$-dim one by a canonical injection:
\bea
R_{3 \times 3} \rightarrow
\left(
\begin {array}{cc}
R_{3 \times 3} & 0\\
0 & 1
\end{array}
\right)
&\Rightarrow& W_{\mbox{\tiny 4ext}} =
\left(
\begin {array}{cc}
W & 0\\
0 & 1
\end{array}
\right),
\eea

As to the extension of $W$, which is a rotation in $SO(3)$ into a $2$-dim matrix, it is carried out by the $1$-to-$2$ homomorphism between $SO(3)$ and its universal covering $SU(2)$, where every rotation will be mapped into an element of $SU(2)$ acting on $2$-dim space. Denoting Pauli matrices by $ {\mbox{\boldmath {$\sigma$}}} $:
\bea
\label{Pauli}
\sigma_1 =
\left(
\begin {array}{cc}
0 & 1\\
1 & 0
\end{array}
\right),\;
\sigma_2 =
\left(
\begin {array}{cc}
0 & -i\\ i & 0
\end{array}
\right),\;
\sigma_3 =
\left(
\begin {array}{cc}
1 & 0\\
0 & 1
\end{array}
\right),
\eea
 we have the following correspondences:
\bea
\label{paulicorrespondence}
R_x \rightarrow \exp{\left(-i  \frac{\t_x}{2}\,\sigma_3\right)},\;\;
R_y \rightarrow \exp{\left(-i  \frac{\t_y}{2}\,\sigma_1\right)},\;\;
R_z \rightarrow \exp{\left(-i  \frac{\t_z}{2}\,\sigma_2\right)},
  \eea
and thus $W$ given by Eq. (\ref{W}) would be extended into the $2 \times 2$ matrix:
\bea
W_{2ext} &=&  \exp{\left(-i  \frac{\pi}{8}\,\sigma_1\right)}\; \exp{\left(-i  \frac{\t_z}{2}\,\sigma_2\right)}\;  \exp{\left(i  \frac{\pi}{8}\,\sigma_1\right)},\nn\\
&=&
\left(
\begin {array}{cc}
c_{z/2} -\displaystyle{\frac{i}{\sqrt{2}}}\; s_{z/2} & -\displaystyle{\frac{1}{\sqrt{2}}}\; s_{z/2}\\
\displaystyle{\frac{1}{\sqrt{2}}}\; s_{z/2} & c_{z/2} +\displaystyle{\frac{i}{\sqrt{2}}}\;  s_{z/2}
\end{array}
\right).
\eea
Thus we have
\bea
\Delta  \stackrel{Z_8}{\longrightarrow} W_{\mbox{\tiny 2ext}}\; \mbox{Diag}\left(\omega^6,\omega^2\right)\; W^\dagger_{\mbox{\tiny 2ext}}\; \Delta
&=& \left(
\begin {array}{cc}-i\, c^2_{z/2} &  -s^2_{z/2} -\displaystyle{\frac{i}{\sqrt{2}}}\; s_z\\
s^2_{z/2} - \displaystyle{\frac{i}{\sqrt{2}}}\; s_z& i\,c^2_{z/2}
\end{array}
\right)
\Delta.
\eea

\item{\bf Charged lepton mass matrix-flavor basis}
The symmetry restrictions in constructing  the charged lepton mass Lagrangian Eq. (\ref{L2}) is similar to what is obtained in the case of ($S^z\times Z_2 \times Z_2'$). The similarity comes from the fact that the charges assigned to the fields  ($L,l^c,\phi$) for the factor $Z_2$ (of $S^z\times Z_2 \times Z_2'$ ) and for $Z_8$ (of $S^z\times Z_8$) are the same. Thus,
 the story repeats itself, and we end up, assuming a hierarchy in the Higgs $\phi$'s fields vevs ($v_3 \gg v_2,v_1$), with a charged lepton mass matrix adjustable to be approximately in the flavor basis. Moreover, we showed in \cite{LCHN2} that the $Z_8$-symmetry forbids the term
$\overline{L}_i\; \phi^\prime_k\;  l^c_j$, and this remains valid in our construction based on $S^z\times Z_8$.
\item{\bf Majorana neutrino mass matrix}

The mass term is generated from the Lagrangian
\bea
 {\cal{L}}_R &=& {1\over 2}\;h^{zk}_{ij}\;\D_k\; \n_{Ri}^T\; C^{-1}\, \n_{Rj}. \label{LR_Z8}
 \eea
Again, comparing with Eq. (\ref{L2}) and doing the substitutions ($\bar{L}_i\rightarrow \n^T_{Ri},\; \phi_j \rightarrow \n_{Rj}$ and $l^c_k \rightarrow \Delta_k$) which should not be taken too much literally but must be considered as a mnemonic device.   Therefore, the story is done over again, in that
 the Yukawa coupling $h^{zk}_{ij}$, ($i,j=1,2,3; k=1,2$) should satisfy the following constraint expressing the invariability of its components:
 \bea
  \label{hzjik}
  S^{z\mbox{\tiny T}}_{{\n_R}\,li}\; h^{zk}_{ij}\; S^z_{\n_R\, jm}\; S^z_{\D kn} &=& h^{zn}_{\,lm},
  \eea
which can be written in an equivalent matrix form similar to Eq. (\ref{fzjikmat}) as
\bea
 \label{hzjikmat}
 S_{\n_R}^{z\mbox{\tiny T}}\;\; {\bf h}^{z \sigma} \; S^z_{\n_R} \;\; (S^z_{\D})_{\sigma \Lambda} &=& {\bf h}^{z\Lambda}.
 \eea where the matrix ${\bf h}^{zj}$ has $h^{zj}_{ik}$  at its $(i,k)$-th entry.

Again, knowing the solution ${\bf h}$ for ``non-rotated'' case we can get the corresponding one for the rotated case as,
\bea
 \label{mathexp}
 {\bf h}^{z\s} &=& W \; {\bf h}^{j} \; W^{\mbox{\tiny T}}\; (W^\dagger_{2ext})_{j \sigma}.
 \eea
We obtained in \cite{LCHN2} considering both the $S$ and $Z_8$ symmetries the following
\bea
\label{hnon-rotated}
{\bf h}^{1}  =
\left(
\begin {array}{ccc}
A_{R1} &0& 0\\
0& 0& 0\\
0&0&0
\end {array}
\right),\;\;
{\bf h}^{2}  =
\left(
\begin {array}{ccc}
0 &0& 0\\
0& C_{R2}& D_{R2}\\
0&D_{R2}&C_{R2}
\end {array}
\right).
\label{csawZ8mass}
\eea
 Making the similarity transformation by $W$:
 \bea
 \label{similarityh}
 {\bf \tilde{h}}^{zk} &=& W \; {\bf h}^k\; W^{\mbox{\tiny T}},
 \eea
 we get the matrices ${\bf \tilde h}^{zk},\; k=1,2$ whose expressions, in terms of new coefficients related to the old coefficients, are given in Appendix C
 (Eq. (\ref{tildeh}) and Eq. (\ref{tildehecoef})). We follow this by the weighted sum ${\bf h}^{z\sigma} = {\bf \tilde{h}}^{zk}\; (W_{2ext}^{\dagger})_{k\sigma}$:
 \bea
 \label{hzmat}
 {\bf h}^{z1} &=& \left(c_{z/2}+\displaystyle{\frac{i}{\sqrt{2}}}\; s_{z/2}\right)\; {\bf \tilde h}^{z1} -\displaystyle{\frac{1}{\sqrt{2}}}\; s_{z/2}\; {\bf \tilde h}^{z2}, \nn\\\nn\\
 {\bf h}^{z2} &=& \displaystyle{\frac{1}{\sqrt{2}}}\; s_{z/2}\; {\bf \tilde h}^{z1} +\left(c_{z/2}-i\,\sqrt{2}\; s_{z/2}\right)\;  {\bf \tilde h}^{z2}.
  \eea
 We can verify explicitly that ${\bf h}^{z1}, {\bf h}^{z2}$ satisfy the requirements of the $S^z \times Z_8$-symmetry:
 \be
 \label{hzjikzmat}
 S_{\n_R}^{z\mbox{\tiny T}}\; {\bf h}^{z \sigma} \; S^z_{\n_R}\; \left(S^z_{\Delta}\right)_{\sigma \Lambda} = {\bf h}^{z\Lambda}.
 \ee

 When the $\Delta$'s acquire vevs ($\Delta^0_{1,2}$), then we get up to leading order in $s_z$
\bea
M^z_R = {\bf h}^{zk}  \Delta^0_k &\approx&
\left(
\begin {array}{ccc}
\Delta^0_1\; A^z_{R1} & 0 & 0\\
0& \Delta^0_2\; C^z_{R2} & \Delta^0_2\; D^z_{R2} \\
0& \Delta^0_2\; D^z_{R2}& \Delta^0_2 \; C^z_{R2}
\end{array}
\right),
\label{MRZ8}
\eea
which is, to leading order, of the form of Eq. (\ref{Mnz}) with $B^z_R=B'^z_R=0,\, C'^z_R=C^z_R$. It is important to stress that the full  expression, without any approximation,
 of the matrix  $M^z_R$ fulfills the form requirement expressed in Eqs. (\ref{Mnz},\ref{Mzncons1},\ref{Mzncons2}). In case of approximating $M^z_R$ up to a certain order in
$s_z$, as is done in Eq. (\ref{MRZ8}), then the relations expressed in Eq. (\ref{Mzncons2}) are still satisfied up to this certain order but there might be violations at the next order.

\item{\bf Dirac neutrino mass matrix}

The Lagrangian responsible for the Dirac neutrino mass matrix is given by Eq. (\ref{L_D}). Following exactly as in the case of $Z_2^2 \times S^z$-symmetry, we find that
Eqs. (\ref{gzjik}, \ref{gzjikmat}) remain valid, and instead of Eq. (\ref{matgexp}) we have the solution as
\bea
\label{matGexp}
{\bf g}^{z \gamma} &=& W \; {\bf g}^{k} \; W^{\mbox{\tiny T}}\; (W_{4ext }^{\mbox{\tiny T}})_{k \gamma}.
\eea
Note that for our extension, we have $W_{4ext}^T = W_{4ext}^\dagger$ and
the corresponding constraint of Eq. (\ref{gzjik}) that fits our case can be written  in matrix form as:
 \bea
 \label{Gzjikmat}
 S_L^{z\mbox{\tiny T}}\; {\bf g}^{z \sigma} \; S^z_{\n_R}\; \left(S^z_{\tilde{\phi^\prime}}\right)_{\sigma \Lambda} = {\bf g}^{z\Lambda}
 \eea

In \cite{LCHN2}, we found the expressions of $g^j_{ik}$ taking into consideration the $S$ and $Z_8$ symmetries:
\be
\label{Gnon-rotated}
\begin{array}{llll}
{\bf g}^{1}  =
\left(
\begin {array}{ccc}
{\cal A}^1 &0& 0\\
0& 0& 0\\
0&0&0
\end {array}
\right), &
{\bf g}^{2}  =
\left(
\begin {array}{ccc}
0 &{\cal B}^2& -{\cal B}^2\\
0& 0& 0\\
0&0&0
\end {array}
\right), &
{\bf g}^{3}  =
\left(
\begin {array}{ccc}
0 &0& 0\\
0& {\cal C}^3& {\cal D}^3\\
0&{\cal D}^3&{\cal C}^3
\end {array}
\right),&
{\bf g}^{4}  =
\left(
\begin {array}{ccc}
0 &{\cal B}^4&{\cal B}^4\\
0& 0& 0\\
0&0&0
\end {array}
\right).
\end{array}
\ee
Applying now the similarity transformations by $W$:
 \be
 \label{similarityG}
 {\bf \tilde{g}}^{zk} = W \; {\bf g}^k\; W^{\mbox{\tiny T}}
 \ee
 we get the matrices ${\bf \tilde g}^{zk}, k=1,2,3,4$ whose expressions, in terms of new coefficients
related to the old ones, are given in Appendix C
 Eq. (\ref{tildeG}) and Eq. (\ref{tildeGecoef}). We follow by the weighted sum ${\bf g}^{z\sigma} = {\bf \tilde{g}}^{zk}\; \left(W_{4ext}^{\mbox{\tiny T}}\right)_{k\sigma}$, and we checked that the obtained ${\bf g}^{z\sigma}, \sigma =1,2,3,4$ do satisfy Eq. (\ref{Gzjikmat}). When the $\phi^\prime$'s get vevs ($v'_k,\, k=1,2,3,4$), the expression of $M^z_D$ turns out
 to be quite complicated, but we have to leading order in $s_z$:
\bea
M^z_D = {\bf g}^{z k}  v'_k &=&\left(
\begin {array}{ccc}
v_1'\, {\cal A}^z_1 & v_2'\, {\cal B}^z_2 + v_4'\, {\cal B}^z_4 & -v_2'\, {\cal B}^z_2 + v_4'\, {\cal B}^z_4\\
0 & v_3'\, {\cal C}^z_3 & v_3'\, {\cal D}^z_3 \\
0 & v_3'\, {\cal D}^z_3 & v_3'\, {\cal C}^z_3
\end{array}
\right),
\label{MD1z}
\eea
which can be matched to the form of Eqs. (\ref{MzpD},\ref{relMzpD}) leading to
\bea
\a = \frac{2v^\prime_4 {\cal B}^z_4}{v^\prime_2 {\cal B}^z_2 - v^\prime_4 {\cal B}^z_4}, && \b =0.
\label{alfafinal}
\eea
If the vevs satisfy $v^\prime_4 \ll v^\prime_2$ and the Yukawa couplings are of the same order, then we get a perturbative  parameter $\a\ll 1$.
This perturbative parameter resurfaces as one perturbative parameter $\chi$ for $M^z_\n$ (Eq. \ref{chixiexp} leading to $\xi=0$) which
was the objective of this section.

\end{itemize}
\section{Discussion and summary}
We determined $S^z$, the $Z_2$ symmetry behind the proposed new rotated $\mu$--$\tau$ neutrino symmetry which leads directly to a pre-given value for $\t_z$.  We showed how the resulting texture can accommodate all the neutrino mass hierarchies. We implemented later the $S^z$-symmetry in the whole lepton sector, and showed how it is able to account for the charged lepton mass hierarchies. We computed, within type-I seesaw, the neutrino mass hierarchies, and showed that $S^z$ can account for enough leptogenesis since it leads exactly to the same results as the symmetry $S$ corresponding to a vanishing $\t_z$.

Whereas invoking perturbations was necessary to amend the experimentally unacceptable vanishing value of $\t_z$, it is still an interesting issue to study the effects of perturbing
$S^z$, at least to adopt, say, another value of $\t_y$ which is predicted by the symmetry to be equal to $\pi/4$. We carry out this study and illustrate the connection between perturbing
$S$ and perturbing $S^z$.  We will report in a future work a complete numerical study contrasting the predictions of the symmetry and its perturbations to experimental data. Nonetheless,
we have found the analytical exact and approximate solutions for the mass spectrum (Eq. \ref{Mspeczp}), mixing and phase angles (Eqs. \ref{1orxyz}, \ref{mix_xyz}, \ref{rhosig} and \ref{delta}) in terms of the unperturbed corresponding quantities (including the $U^{0z}$ parameters $\t_{z0},\varphi$ and $\xi$ of Eq. \ref{Uzgeneral}) and the perturbation parameters $\a_{ij}$ (Eq. \ref{pertMn}), which can
originate from $\a$, a perturbing parameter in the Dirac mass matrix (Eq. \ref{pertMD}). Surely, one needs a phenomenological test to give an order of magnitude for
$\a$ and show how it is reflected in the full mass spectrum, mixing and phase angles. One needs also to perform a thorough numerical analysis scan over the mass matrix perturbation parameters
 ($\a_{ij}$) in order to seek viable choices. Although this numerical scan goes beyond the scope of this paper, however one can argue roughly that as we have carried out in \cite{LCHN1,LCHN2} a complete numerical analysis of perturbing the $S$ symmetry (which was necessary to move $\t_z$ from $0^o$ to $\sim 10^o$), then one can take the same similarity rotation $W$ and apply it to
 the phenomenologically viable perturbation $\d M_\n$ to get (cf. Eq. \ref{cond}):
 \bea
 \label{argument1}
 \d M_\n^z = W\, \d M_\n\, W^T &\Rightarrow& M_\n^z = W\, M_\n\, W^T, \\
 &&
 \label{argument2}
 U^z= W\,U=W\,U^0\,\left(1+I_\eps\right)=W\,U^0+W\,U^0\,I_\eps.
 \eea
 Now, for the first implication (Eq. \ref{argument1}), it means that the perturbed mass spectrum of the $S^z$ case is identical to the phenomenologically viable
 perturbed mass spectrum in the $S$ case for the ``successful'' choices in its scan. As to the second implication concerning the diagonalizing matrix (Eq. \ref{argument2}), the first term $WU^0$, corresponding to rotated non-perturbed texture, would lead to $\t_y=\pi/4, \t_z=\t_{z0}$. Since the second term $WU^0I_\eps$ remains infinitesimal, then it would not change much the mixing angles and we expect
 that phenomenological viable regions in the parameter space to exist.

Finally, we presented a theoretical realization of the perturbed Dirac mass matrix, where the symmetry is broken spontaneously and the perturbation parameter originates from ratios of different Higgs fields vevs.

\section*{{\large \bf Acknowledgements}}
E.I.L. thanks the INFN (Padova section), where some part of this
work has been done, for its hospitality.
 N.C. acknowledges funding provided by the Alexander von Humboldt Foundation. This project has received funding from the European Union's Horizon 2020 research and innovation programme under the Marie Sklodowska-Curie grant agreement No 690575.


\appendix
\section{The adopted parametrization for $M_\n$ and constraints on $M_\n$, $M_R$ and $M_D$ from the $S$ and $S^z$ symmetries}
\begin{itemize}
\item
In the flavor basis and in the parametrization adopted in our work, the elements of the neutrino mass matrix are given by:
\bea
M_{\n\,11}&=& m_1\, c_{x}^2\, c_{z}^2\, e^{2\,i\,\r} + m_2\, s_{x}^2\, c_{z}^2\, e^{2\,i\,\s}
+ m_3\,s_{z}^2,\nn\\
M_{\n\,12}&=& m_1\,\left( - c_{z}\, s_{z}\, c_{x}^2 \,s_{y} e^{2\,i\,\r}
- c_{z}\, c_{x} s_{x}\, c_{y}\, e^{i\,(2\,\r-\d)}\right)\nn \\ &&
+ m_2\,\left( - c_{z}\, s_{z}\, s_{x}^2\, s_{y} e^{2\,i\,\s}
+ c_{z}\, c_{x}\, s_{x}\, c_{y}\, e^{i\,(2\,\s-\d)}\right) + m_3\, c_{z}\, s_{z}\, s_{y},\nn\\
M_{\n\,13}&=& m_1\,\left( - c_{z}\, s_{z}\, c_{x}^2\, c_{y}\, e^{2\,i\,\r}
+ c_{z}\, c_{x}\, s_{x}\, s_{y}\, e^{i\,(2\,\r-\d)}\right)\nn \\ &&
+ m_2\,\left( - c_{z}\, s_{z}\, s_{x}^2\, c_{y}\, e^{2\,i\,\s}
- c_{z}\, c_{x}\, s_{x}\, s_{y}\, e^{i\,(2\,\s-\d)}\right) + m_3\, c_{z}\, s_{z}\, c_{y},\nn\\
M_{\n\,22}&=& m_1\, \left( c_{x}\, s_{z}\, s_{y}\,  e^{i\,\r}
+ c_{y}\, s_{x}\, e^{i\,(\r-\d)}\right)^2\nn \\ &&
 + m_2\, \left( s_{x}\, s_{z}\, s_{y}\,  e^{i\,\s}
- c_{y}\, c_{x}\, e^{i\,(\s-\d)}\right)^2 + m_3\, c_{z}^2\, s_{y}^2, \nn\\
M_{\n\,33}&=& m_1\, \left( c_{x}\, s_{z}\, c_{y}\,  e^{i\,\r}
- s_{y}\, s_{x}\, e^{i\,(\r-\d)}\right)^2\nn \\ &&
 + m_2\, \left( s_{x}\, s_{z}\, c_{y}\,  e^{i\,\s}
+ s_{y}\, c_{x}\, e^{i\,(\s-\d)}\right)^2 + m_3\, c_{z}^2\, c_{y}^2, \nn\\
M_{\n\, 23} &=& m_1\,\left( c_{x}^2\, c_{y}\, s_{y}\, s_{z}^2\,  e^{2\,i\,\r}
 + s_{z}\, c_{x}\, s_{x}\, \left(c_{y}^2-s_{y}^2\right)\, e^{i\,(2\,\r-\d)} - c_{y}\, s_{y}\, s_{x}^2\, e^{2\,i\,(\r-\d)}\right)
\nn\\
&&
+  m_2\,\left( s_{x}^2\, c_{y}\, s_{y}\, s_{z}^2\,  e^{2\,i\,\s}
 + s_{z}\, c_{x}\, s_{x}\, \left(s_{y}^2-c_{y}^2\right)\, e^{i\,(2\,\s-\d)} - c_{y}\, s_{y}\, c_{x}^2\, e^{2\,i\,(\s-\d)}\right)\nn \\ &&
+ m_3\, s_{y}\, c_{y}\, c_{z}^2.
\label{melements}
\eea

\item
The invariance of the symmetric $M_\n$ under the symmetry $S$ implies the following forms:
\bea
M_{\nu} = \left(\begin{array}{ccc}
A_{\nu} & B_{\nu} &  -B_{\nu} \\
B_{\nu} & C_{\nu}  & D_{\nu} \\
-B_{\nu} & D_{\nu} & C_{\nu} \end{array}\right) &,& M_{\nu}^* \,M_{\nu} =
 \left(\begin{array}{ccc}
 a_\n & b_\n &  - b_\n \\
b_\n^* &  c_\n & d_\n\\
-b_\n^* & d_\n & c_\n
\end{array}
\right),
\label{MnS}
\eea
where
\bea
  a_\n  =  \left|A_\n\right|^2 + 2 \left|B_\n\right|^2, && b_\n  = A_\n^*\,B_\n + B_\n^*\,C_\n - B_\n^*\,D_\n,\nn \\
  c_\n  =  \left|B_\n\right|^2 +  \left|C_\n\right|^2 + \left|D_\n\right|^2, &&
  d_\n  =  -\left|B_\n\right|^2 + C_\n^*\,D_\n + D_\n^*\,C_\n.
  \label{abs+}
  \eea
The mass spectrum for $M_{\nu}$ and $M_{\nu}^* \,M_{\nu}$ are respectively given by:
\bea
M_{\n\,11}^{\mbox{\tiny Diag}} &=& A_\n\,c_\varphi^2 - \sqrt{2}\,s_{2\varphi}\,e^{-i\,\xi}\,B_\n + \left(C_\n - D_\n\right)\,s_\varphi^2\,e^{-2\,i\,\xi},\nn\\
M_{\n\,22}^{\mbox{\tiny Diag}} &=& A_\n\,s_\varphi^2 + \sqrt{2}\,s_{2\varphi}\,e^{-i\,\xi}\,B_\n + \left(C_\n - D_\n\right)\,c_\varphi^2\,e^{-2\,i\,\xi},\nn\\
M_{\n\,33}^{\mbox{\tiny Diag}} &=& C_\n + D_\n,
\label{Mdiags+}
\eea
and
\bea
m_1^2 & = & {a_\n + c_\n - d_\n\over 2} + {1\over 2} \sqrt{\left(a_\n+d_\n-c_\n\right)^2 + 8\, \left|b_\n\right|^2},\nn\\
 m_2^2  & = & {a_\n + c_\n - d_\n\over 2} - {1\over 2} \sqrt{\left(a_\n+d_\n-c_\n\right)^2 + 8\, \left|b_\n\right|^2},\nn\\
m_3^2 & = & c_\n + d_\n.
\label{msqs+}
\eea

\item
The invariance of the symmetric $M^z_\n$ under the symmetry $S^z$ implies that $M^z_\n$ can be written in terms of four independent complex parameters $\left\{A_\n^z, B_\n^z, C_\n^z, D_\n^z\right\}$ as
\bea
M_{\n\,11}^z &=& A_\n^z,\;\;  M_{\n\,12}^z = B_\n^z,\;\;
M_{\n\,13}^z = -\left(1-2\,t_z^2\right)\,B^z_\n - \sqrt{2} t_z\,\left(A^{z}_\n+ C^z_\n -D^z_\n\right),  \nn \\
M_{\n\,21}^z &=& M_{\n\,12}^z,\;\;  M_{\n\,22}^z = C_\n^z,\;\;  M_{\n\,23}^z = D_\n^z, \nn \\
M_{\n\,31}^z & = & M_{\n\,13}^z,\;\;
 M_{\n\,32}^z  = M_{D\,23}^z,\;\;
 M_{\n\,33}^z  = \left(1-2t_z^2\right) C_\n^{z}  + 2\, t_z^2\, \left(A_\n^{z} - D_\n^{z}\right) + 2\,\sqrt{2}\, t_z\, \left(1- t_z^2\right)\, B_\n^{z},
 \label{Mzncon3}
 \eea
where the set of parameters $\left\{A_\n^{z}, B_\n^{z}, C_\n^{z}, D_\n^{z}\right\}$ can be written in terms of the set $\left\{A_\n, B_\n, C_\n, D_\n\right\}$ as,
\bea
\label{relAnzAn}
A^z_\n &=& c_z^2 A_\n + s_z^2 \left(D_\n+C_\n\right), \nonumber \\
B^z_\n &=& c_z B_\n -\displaystyle{{s_{2z}\over 2\sqrt{2}}}\, \left(A_\n-C_\n-D_\n\right), \nonumber \\
C^z_\n &=& \displaystyle{{1\over 2}}\, \left(1+c_z^2\right) C_\n -\sqrt{2} s_z B_\n + \displaystyle{{s_{z}^2\over 2}}\, \left(A_\n-D_\n\right), \nonumber \\
 D^z_\n &=& \displaystyle{{1\over 2}}\, \left(1+c_z^2\right)\, D_\n + \displaystyle{{s_z^2\over 2}}\,\left(A_\n-C_\n\right).
 \eea
 The inverse relations can be expressed as,
 \bea
\label{relAnAnz}
A_\n &=& \displaystyle{{1\over c_z^2}} A^z_\n - \sqrt{2}\,t_z^3\, B^z_\n - t_z^2\, \left(C_\n^z + D_\n^z\right), \nonumber \\
B_\n &=& \displaystyle{{c_{2z}\over c_z^3}}\, B_\n^z +\displaystyle{{t_z\over \sqrt{2}\,c_z}} \left(A_\n^z-C_\n^z-D_\n^z\right), \nonumber \\
C_\n &=& \left(1- \displaystyle{{t_z^2\over 2}}\right) C_\n^z  + \displaystyle{{t_{z}^2\over 2}}\, \left(A_\n^z-D_\n^z\right) + \sqrt{2}\,t_z\,\left(1- \displaystyle{{t_z^2\over 2}}\right)\,B_\n^z, \nonumber \\
 D_\n &=& \left(1 + \displaystyle{{t_z^2\over 2}}\right)\, D_\n^z + \displaystyle{{t_z^3\over \sqrt{2}}}\, B_\n^z - \displaystyle{{t_z^2\over 2}}\,\left(A_\n^z-C_\n^z\right).
 \eea

 As to the hermitian mass squared matrix $M_\n^{z*}M_\n^z$, it should have the form:
 \bea \label{mz2Par}
(M_\n^{z*}M_\n^z)_{11} &=& a_\n\, c_z^2 + s_z^2\, \left(c_\n +d_\n\right), \nonumber \\
(M_\n^{z*}M_\n^z)_{12} &=& b_\n\, c_z -\displaystyle{{s_{2z}\over 2\sqrt{2}}}\, \left(a_\n -c_\n -d_\n\right), \nonumber \\
(M_\n^{z*}M_\n^z)_{13} &=& - b_\n\, c_z + \displaystyle{{s_{2z}\over 2\sqrt{2}}}\, \left(a_\n -c_\n -d_\n\right), \nonumber \\
(M_\n^{z*}M_\n^z)_{21} &=& b_\n^*\,c_z -\displaystyle{{s_{2z}\over 2\sqrt{2}}}\, \left(a_\n -c_\n -d_\n\right), \nonumber \\
(M_\n^{z*}M_\n^z)_{22} &=& \displaystyle{1\over 2}\,\left(1+c_z^2\right)\, c_\n + \displaystyle{{s_z^2\over 2}}\, \left(a_\n -d_\n\right)  -\sqrt{2}\, \mbox{Re}\left(b_\n\right) s_z, \nonumber \\
(M_\n^{z*}M_\n^z)_{23} &=& \displaystyle{1\over 2}\, \left(1+c_z^2\right) d_\n + \displaystyle{{s_z^2\over 2}}\, (a_\n -c_\n)  + i \sqrt{2}\,  \mbox{Im}(b_\n) s_z \nonumber \\
(M_\n^{z*}M_\n^z)_{31} &=& - b_\n^*c_z -\displaystyle{{s_{2z}\over2\sqrt{2}}}\, \left(a_\n -c_\n -d_\n\right), \nonumber \\
(M_\n^{z*}M_\n^z)_{32} &=& \displaystyle{{1\over 2}}\,\left(1+c_z^2\right)\,d_\n + \displaystyle{{s_z^2\over 2}}\, \left(a_\n -c_\n\right)  - i\, \sqrt{2}\,  \mbox{Im}\left(b_\n\right)\, s_z, \nonumber \\
(M_\n^{z*}M_\n^z)_{33} &=& \displaystyle{{1\over 2}}\, \left(1+c_z^2\right)\, c_\n + \displaystyle{{s_z^2\over 2}}\, \left(a_\n -d_\n\right)  +\sqrt{2}\, \mbox{Re}\,\left(b_\n\right)\, s_z.
\eea

\item
All results derived for $M_\n$ and $M_\n^z$ concerning symmetry properties under $S$ and $S^z$ and their mutual interrelations would equally apply to the case of $M_R$ and $M_R^z$. More precisely, we have all the formulae from Eq. (\ref{MnS}) till Eq. (\ref{mz2Par}) with the appropriate replacement of the subscript $\n$ into $R$.

\item
The invariance of the Dirac neutrino mass matrix $M_D$ under the symmetry $S$ implies the following forms:
\bea
M_D = \left(
\begin{array}{ccc}
A_D & B_D &  -B_D \\
E_D & C_D  & D_D \\
-E_D & D_D & C_D
\end{array}
\right)
&,& M_D^\dagger \,M_D =
 \left(\begin{array}{ccc}
 a_D & b_D &  -b_D \\
b_D^* &  c_D & d_D\\
-b_D^* & d_D & c_D
\end{array}\right),
\label{MD1}
\eea
where
\bea
  a_D  =  \left|A_D\right|^2 + 2 \left|E_D\right|^2, && b_D  = A_D^*\,B_D + E_D^*\,C_D - E_D^*\,D_D,\nn \\
  c_D  =  \left|B_D\right|^2 +  \left|C_D\right|^2 + \left|D_D\right|^2, &&
  d_D  =  -\left|B_D\right|^2 + C_D^*\,D_D + D_D^*\,C_D.
  \label{MD2}
  \eea
The mass spectrum of $M_D^\dagger \,M_D$ can be written as
\be
\left\{\; c_{D} + d_{D},\; {a_{D} + c_{D} - d_{D}\over 2} \pm {1\over 2} \sqrt{\left(a_{D}+d_{D}-c_{D}\right)^2 + 8\, \left|b_{D}\right|^2}
\;\right\}.
\label{specMD}
\ee

\item
The invariance of $M^z_D$ under the symmetry $S^z$ implies that $M^z_D$ can be written in terms of five independent complex parameters $\left\{A_D^z, B_D^z, C_D^z, D_D^z, E_D^z\right\}$ as
\bea
M_{D\,11}^z &=& A_D^z,\;\;  M_{D\,12}^z = B_D^z,\;\;
M_{D\,13}^z = -B^{z}_D + \sqrt{2} t_z\,\left(D^{z}_D+ C^{z}_D -A^{z}_D\right) + 2 t_z^2 E^{z}_D,  \nn \\
M_{D\,21}^z &=& E_D^z,\;\;  M_{D\,22}^z = C_D^z,\;\;  M_{D\,23}^z = D_D^z, \nn \\
M_{D\,31}^z & = & -\left(1-2t_z^2\right) E_D^{z} + \sqrt{2} t_z\, \left(D_D^{z}+C_D^{z}-A_D^{z}\right),\;\;
 M_{D\,32}^z  = D_D^{z} + \sqrt{2} t_z\, \left(E^{z}_D-B^{z}_D\right), \nn  \\
 M_{D\,33}^z  &=& \left(1-2t_z^2\right) C_D^{z} + \sqrt{2}\, t_z\, B_D^{z} + 2\, t_z^2\, \left(A_D^{z} - D_D^{z}\right) + \sqrt{2}\, t_z\, \left(1-2 t_z^2\right)\, E_D^{z}.
 \label{Mzdcon3}
 \eea
The set of parameter $\left\{A_D^z, B_D^z, C_D^z, D_D^z, E_D^z\right\}$  can be written in terms of $\left\{A_D, B_D, C_D, D_D, E_D\right\}$ as
\bea
\label{relADzAD}
A^z_D &=& c_z^2 A_D + s_z^2 \left(D_D + C_D\right), \nonumber \\
B^z_D &=& c_z B_D -\displaystyle{{s_{2z}\over 2\sqrt{2}}}\, \left(A_D-C_D-D_D\right), \nonumber \\
C^z_D &=& \displaystyle{{1\over 2}}\, \left(1+c_z^2\right) C_D + \displaystyle{{s_z^2\over 2}} \,\left(A_D - D_D\right) - \displaystyle{{s_z\over \sqrt{2}}}\, \left(B_D + E_D\right), \nonumber \\
D^z_D &=& \displaystyle{{1\over 2}}\, \left(1+c_z^2\right) D_D +\displaystyle{{s_z^2\over 2}} \,\left(A_D - D_D\right) - \displaystyle{{s_z\over \sqrt{2}}}\, \left(B_D - E_D\right), \nonumber \\
 E_D^z &=& c_z\, E_D - \displaystyle{{s_{2z}\over 2\,\sqrt{2}}}\,\left(A_D- D_D - C_D\right),
 \eea
 while the inverse relations can written as
 \bea
\label{relADADz}
A_D &=& \left(1 + t_z^2\right)\,A_D^z -\sqrt{2}\,t_z^3\, E_D^z - t_z^2\,\left(C_D^z + D_D^z\right), \nonumber \\
B_D &=& \displaystyle{{1\over c_z}}\, B_D^z + \displaystyle{{t_z\over \sqrt{2} c_z}}\,\left(A_D^z - C_D^z - D_D^z\right) - \displaystyle{{t_z^2 \over  c_z}}\,E_D^z, \nonumber \\
C_D &=& \left(1- \displaystyle{{t_z^2\over 2}}\right) C_D^z - \displaystyle{{t_z^2\over 2}}\,\left(D_D^z - A_D^z \right) + \displaystyle{{t_z\over \sqrt{2}}}\,B_D^z + \displaystyle{{1\over \sqrt{2}}}\,t_z\,\left(1-t_z^2\right)\, E_D^z, \nonumber \\
D_D &=& \left(1+ \displaystyle{{t_z^2\over 2}}\right) D_D^z + \displaystyle{{t_z^2\over 2}}\,\left(C_D^z - A_D^z \right) - \displaystyle{{t_z\over \sqrt{2}}}\,B_D^z + \displaystyle{{t_z\over \sqrt{2}\,c_z^2}}\, E_D^z, \nonumber \\
 E_D &=& \displaystyle{{1\over c_z}}\,\left(1- t_z^2\right) E_D^z + \displaystyle{{t_z\over \sqrt{2}\, c_z}}\,\left(A_D^z - C_D^z - D_D^z\right).
 \eea

 As to the hermitian mass squared matrix $M_D^{z*}M_D^z$, it should have the form:
 \bea
 \label{MDz2Par}
(M_D^{z\dagger}M_D^z)_{11} &=& a_D\, c_z^2 + s_z^2\, \left(c_D +d_D\right), \nonumber \\
(M_D^{z\dagger}M_D^z)_{12} &=& b_D\, c_z -\displaystyle{{s_{2z}\over 2\sqrt{2}}}\, \left(a_D -c_D -d_D\right), \nonumber \\
(M_D^{z\dagger}M_D^z)_{13} &=& - b_D\, c_z + \displaystyle{{s_{2z}\over 2\sqrt{2}}}\, \left(a_D -c_D -d_D\right), \nonumber \\
(M_D^{z\dagger}M_D^z)_{21} &=& b_D^*\,c_z -\displaystyle{{s_{2z}\over 2\sqrt{2}}}\, \left(a_D -c_D -d_D\right), \nonumber \\
(M_D^{z\dagger}M_D^z)_{22} &=& \displaystyle{1\over 2}\,\left(1+c_z^2\right)\, c_D + \displaystyle{{s_z^2\over 2}}\, \left(a_D -d_D\right)  -\sqrt{2}\, \mbox{Re}\left(b_D\right) s_z, \nonumber \\
(M_D^{z\dagger}M_D^z)_{23} &=& \displaystyle{1\over 2}\, \left(1+c_z^2\right) d_D + \displaystyle{{s_z^2\over 2}}\, (a_D -c_D)  + i \sqrt{2}\,  \mbox{Im}(b_D) s_z \nonumber \\
(M_D^{z\dagger}M_D^z)_{31} &=& - b_D^*c_z -\displaystyle{{s_{2z}\over2\sqrt{2}}}\, \left(a_D -c_D -d_D\right), \nonumber \\
(M_D^{z\dagger}M_D^z)_{32} &=& \displaystyle{{1\over 2}}\,\left(1+c_z^2\right)\,d_D + \displaystyle{{s_z^2\over 2}}\, \left(a_D -c_D\right)  - i\, \sqrt{2}\,  \mbox{Im}\left(b_D\right)\, s_z, \nonumber \\
(M_D^{z\dagger}M_D^z)_{33} &=& \displaystyle{{1\over 2}}\, \left(1+c_z^2\right)\, c_D + \displaystyle{{s_z^2\over 2}}\, \left(a_D -d_D\right)  +\sqrt{2}\, \mbox{Re}\,\left(b_D\right)\, s_z.
\eea

\item
The invariance of generic symmetric matrix $\left(M=M^T\right)$ under the $S^z$-sign-flipped symmetry  defined as
\be
S^{z\mbox{\textsc{t}}} \, M \, S^z=-M,
\label{invsz}
\ee
 leads to the fact that the symmetric matrix  $M$ can be written in terms of two independent complex parameters $\left\{B,\, C\right\}$ as,
\bea
&&M_{11} = 2\,\sqrt{2}\, t_z\, B - 2\,t_z^2\, C,\; M_{12} = B,\; M_{13} = \left(1-2\,t_z^2\right)\,B -\sqrt{2}\, t_z\, \left(1-t_z^2\right)\, C\nn\\
&&M_{22} = C,\; M_{23} = -\sqrt{2}\, t_z\, B + t_z^2\, C, \; M_{33} = -2\, \sqrt{2}\, t_z\, B - \left(1-2\,t_z^2\right)\, C.
\label{invszsymM}
\eea

The invariance as represented in Eq. (\ref{invsz}) but for generic mass matrix $M$ implies that $M$ can be written in terms of four independent complex parameters $\left\{B,\, C,\, D,\, E\right\}$ as,
\bea
&&M_{11} = \sqrt{2}\, t_z\, \left(B+E\right) - 2\,t_z^2\, C,\; M_{12} =  B, \; M_{13} = B + \sqrt{2}\, t_z\, \left(D - C\right),\nn\\
&& M_{21} = E,\; M_{22} = C,\; M_{23} = D,\; M_{31} = \left(1-2\,t_z^2\right)\, E -2\, t_z^2\, B - \sqrt{2}\,t_z\, \left(1-2\,t_z^2\right)\, C -\sqrt{2}\, t_z\, D,\nn\\
&&
M_{32} = -D -\sqrt{2}\, t_z\, \left(B+E\right) + 2\, t_z^2\, C,\;
M_{33}= -\sqrt{2}\,t_z\,\left(B+E\right) - \left(1-2\,t_z^2\right)\, C.
\label{invszgenM}
\eea
\end{itemize}

\section{Perturbing $S^z$}
\begin{itemize}
\item
For a proof that the two determinations of $I_\eps^z$ through Eq. (\ref{condIepsz}) and Eq. (\ref{condIepsz2}) are equivalent, we start from the following two equations that force the diagonalization of $M_\n^z$, namely,
\be
\label{step1}
\begin{array}{lll}
\left[-I^{z*}_\eps\, \,M_\nu^{0z\mbox{\tiny Diag}}  +  M_\nu^{0z\mbox{\tiny Diag}}\, I^{z}_\eps + U^{0z\mbox{\tiny T}}\, \d M^z_\n\, U^{0z} \right]_{i\,\neq\,j} = 0, &&\nn\\\\
\left[-I^{z}_\eps\, \,M_\nu^{0z*\mbox{\tiny Diag}}  +  M_\nu^{0z*\mbox{\tiny Diag}} I^{z*}_\eps + U^{0z*\mbox{\tiny T}} \d M^{z*}_\n U^{0z*} \right]_{i\,\neq\,j} = 0.&&
\end{array}
\ee
Multiplying the first one from left by $ M_\nu^{0z\mbox{\tiny Diag}*}$ while the second from  right by $ M_\nu^{0z\mbox{\tiny Diag}}$, and then adding the resulting two equations we get
\bea
\label{step2}
\left(\left[M_\nu^{0z\mbox{\tiny Diag}*}\,M_\nu^{0z\mbox{\tiny Diag}},\;I^z_\epsilon\right] +
M_\nu^{0z\mbox{\tiny Diag}*}\;U^{0z\,T}\; \d M^{z}_\n\;U^{0z} +
U^{0z\,*T}\;\d M^{z*}_\n\;U^{0z\,*T}\; M_\nu^{0z\mbox{\tiny Diag}}\right)_{i\,\neq\,j} =0. &&
\eea
Upon using Eq. (\ref{U0z1}), the above equation gets transformed into,
\bea
\label{step3}
\left(\left[M_\nu^{0z\mbox{\tiny Diag}*}\,M_\nu^{0z\mbox{\tiny Diag}},\;I^z_\epsilon\right] +
U^{0z\,\dagger}\;M_\nu^{0z*}\; \d M^{z}_\n\;U^{0z} +
U^{0z\,\dagger}\;\d M^{z*}_\n\; M_\nu^{0z}\;U^{0z}\right)_{i\,\neq\,j} =0. &&
\eea
The above equation is nothing but the one that forces the diagonalization of the combination $M_\n^{z*}\, M_\n^z$ as written in Eq. (\ref{condIepsz}). Conversely, one can easily go backward which means that Eq. (\ref{step3}) leads to Eq. (\ref{step1}).

\item
Solving the linear system of Eqs. (\ref{condIepsz2})  we get:
\bea
\label{epsilon}
\eps_1^z &=& \displaystyle{{1\over \left|M_{\n\,11}^{0z\tiny{\mbox{Diag}}}\right|^2-
\left|M_{\n\,22}^{0z\tiny{\mbox{Diag}}}\right|^2}}\,
\left\{\displaystyle{{1\over\sqrt{2}}}\, c_{2\varphi}\, c_{z0}\, M_{\n\,11}^{0z\tiny{\mbox{Diag}}*}\,\left(\a_{13}^z-\a_{12}^z \right)\,e^{-i\xi} + \displaystyle{{1\over \sqrt{2}}}\, c_{2\varphi}\, c_{z0} M_{\n\,22}^{0z\tiny{\mbox{Diag}}}\,\left(\a^{z*}_{13}-\a^{z*}_{12}\right)\,e^{i\xi} \right. \nonumber \\ \nn\\&&\!\!\!\!
\!\!\!\!\left. -M_{\n\,22}^{0z\tiny{\mbox{Diag}}}\, c_{z0}\, s_{\varphi}\, c_{\varphi} \left[-\sqrt{2}\,s_{z0}\, \left(\a_{12}^{z*} + \a_{13}^{z*}\right)+c_{z0} \a_{11}^{z*}\right] -M_{\n\,11}^{0z\tiny{\mbox{Diag}}*}\, c_{z0} s_{\varphi}\, c_{\varphi}\, \left[-\sqrt{2}\,s_{z0}\, \left(\a_{12}^z + \a_{13}^z\right)+ c_{z0}\, \a_{11}^z\right] \right\},
\nonumber \\ \nn\\
 \eps_2^z &=& \displaystyle{{1\over \left|M_{\n\,11}^{0z\tiny{\mbox{Diag}}}\right|^2-
\left|M_{\n\,33}^{0z\tiny{\mbox{Diag}}}\right|^2}}\,
\left\{-\displaystyle{{1\over\sqrt{2}}}\, s_{\varphi}\, s_{z0}\, M_{\n\,11}^{0z\tiny{\mbox{Diag}}*}\,\left(\a_{13}^z-\a_{12}^z \right)\,e^{-i\xi} - \displaystyle{{1\over \sqrt{2}}}\, s_{\varphi}\, s_{z0} M_{\n\,33}^{0z\tiny{\mbox{Diag}}}\,\left(\a^{z*}_{13}-\a^{z*}_{12}\right)\,e^{i\xi} \right. \nonumber \\ \nn\\&&
\left. -\displaystyle{{1\over 2}}\,M_{\n\,33}^{0z\tiny{\mbox{Diag}}}\, c_{\varphi} \left[\sqrt{2}\,c_{2z0}\, \left(\a_{12}^{z*} + \a_{13}^{z*}\right)+s_{2z0}\, \a_{11}^{z*}\right] - \displaystyle{{1\over 2}}\,M_{\n\,11}^{0z\tiny{\mbox{Diag}}*}\, c_{\varphi}\, \left[\sqrt{2}\,c_{2z0}\, \left(\a_{12}^z + \a_{13}^z\right)+ s_{2z0}\, \a_{11}^z\right] \right\},
\nonumber \\ \nn\\
\eps_3^z &=& \displaystyle{{1\over \left|M_{\n\,22}^{0z\tiny{\mbox{Diag}}}\right|^2-
\left|M_{\n\,33}^{0z\tiny{\mbox{Diag}}}\right|^2}}\,
\left\{\displaystyle{{1\over\sqrt{2}}}\, c_{\varphi}\, s_{z0}\, M_{\n\,22}^{0z\tiny{\mbox{Diag}}*}\,\left(\a_{13}^z-\a_{12}^z \right)\,e^{-i\xi} + \displaystyle{{1\over \sqrt{2}}}\, c_{\varphi}\, s_{z0} M_{\n\,33}^{0z\tiny{\mbox{Diag}}}\,\left(\a^{z*}_{13}-\a^{z*}_{12}\right)\,e^{i\xi} \right. \nonumber \\ \nn\\&&
\left. -\displaystyle{{1\over 2}}\,M_{\n\,33}^{0z\tiny{\mbox{Diag}}}\, s_{\varphi} \left[\sqrt{2}\,c_{2z0}\, \left(\a_{12}^{z*} + \a_{13}^{z*}\right)+s_{2z0}\, \a_{11}^{z*}\right] - \displaystyle{{1\over 2}}\,M_{\n\,22}^{0z\tiny{\mbox{Diag}}*}\, s_{\varphi}\, \left[\sqrt{2}\,c_{2z0}\, \left(\a_{12}^z + \a_{13}^z\right)+ s_{2z0}\, \a_{11}^z\right] \right\}.\nonumber\\
\eea

The complete entries of the matrix $U_\eps$ are
\bea
U_\eps\left(1,1\right) &=& \left(c_{z0}\, c_\varphi - c_{z0}\, s_\varphi\, \eps_1^{z*} - s_{z0}\, \eps_2^{z*}\right)\, e^{-i\,\phi_1},\nn\\
U_\eps\left(1,2\right) &=& \left(c_{z0}\, s_\varphi + c_{z0}\, c_\varphi\, \eps^z_1 - s_{z0}\, \eps_3^{z*}\right)\, e^{-i\,\phi_2},\nn\\
U_\eps\left(1,3\right) &=& \left(s_{z0}\, + c_{z0}\, c_\varphi\, \eps^z_2 + c_{z0}\, s_\varphi \eps^z_3\right)\, e^{-i\,\phi_3},\nn\\
U_\eps\left(2,1\right) &=&-\displaystyle{{1\over \sqrt{2}}} \left(s_{z0}\,c_\varphi +  s_\varphi\, e^{-i\,\xi} - s_{z0}\,s_\varphi \eps_1^{z*} + c_\varphi\, \eps_1^{z*} e^{-i\,\xi} + c_{z0}\, \eps_2^{z*} \right)\, e^{-i\,\phi_1},\nn\\
U_\eps\left(2,2\right) &=&-\displaystyle{{1\over \sqrt{2}}} \left(s_{z0}\,s_\varphi -  c_\varphi\, e^{-i\,\xi} + s_{z0}\,c_\varphi \eps^z_1 + s_\varphi\, \eps^z_1\, e^{-i\,\xi} + c_{z0}\, \eps_3^{z*} \right)\, e^{-i\,\phi_2},\nn\\
U_\eps\left(2,3\right) &=&\displaystyle{{1\over \sqrt{2}}} \left(c_{z0} -  c_\varphi\,s_{z0} \, \eps^z_2 - s_\varphi\, \eps^z_2\,e^{-i\,\xi} -  s_{z0}\, s_\varphi\, \eps^z_3 +  c_\varphi\, \eps^z_3\, e^{-i\,\xi} \right)\, e^{-i\,\phi_3},\nn\\
U_\eps\left(3,1\right) &=&-\displaystyle{{1\over \sqrt{2}}} \left(s_{z0}\, c_\varphi -  s_\varphi \,e^{-i\,\xi}  - s_{z0} s_\varphi\, \eps_1^{z*}  -   c_\varphi\, \eps_1^{z*}\,\,e^{-i\,\xi} +  c_{z0} \, \eps_2^{z*} \right)\, e^{-i\,\phi_1},\nn\\
U_\eps\left(3,2\right) &=&-\displaystyle{{1\over \sqrt{2}}} \left(s_{z0}\, s_\varphi +  c_\varphi \,e^{-i\,\xi}  + s_{z0} c_\varphi\, \eps^z_1  -   s_\varphi\, \eps^z_1\,\,e^{-i\,\xi} +  c_{z0} \, \eps_3^{z*} \right)\, e^{-i\,\phi_2},\nn\\
U_\eps\left(3,3\right) &=&\displaystyle{{1\over \sqrt{2}}} \left(c_{z0} -  c_\varphi s_{z0}\, \eps^z_2   +  s_\varphi\, \eps^z_2\,e^{-i\,\xi}  -   s_{z0}\,s_\varphi\, \eps^z_3 +  c_\varphi \, \eps^z_3\, e^{-i\,\xi} \right)\, e^{-i\,\phi_3}.
\label{uepsent}
\eea

The corresponding mixing angles are the following.
\bea
\label{mix_xyz}
t_x = \left|\displaystyle{{U_\eps\left(1,2\right)\over U_\eps\left(1,1\right)}}\right| &=&  \displaystyle{{\left|c_{z0}\,  s_{\varphi} + c_{z0}\,  c_{\varphi}\,  \eps^z_1  - s_{z0}\, \eps_3^{z*}  \over c_{z0}\,  c_{\varphi} - c_{z0}\,  s_{\varphi}  \eps_1^{z*}      - s_{z0}\, \eps_2^{z*}\right|}},\nn\\ \nn\\
t_y = \left|\displaystyle{{U_\eps\left(2,3\right)\over U_\eps\left(3,3\right)}}\right|  &=& \displaystyle{{\left|c_{z0}-\eps^z_2\, s_{z0}\, s_{\varphi} - \eps^z_2\, s_\varphi\, e^{-i\xi} -  \eps^z_3\, s_{z0}\, s_{\varphi} + \eps^z_3\, c_\varphi\, e^{-i\xi} \over c_{z0}-\eps^z_2\, s_{z0}\, c_{\varphi} + \eps^z_2\, s_\varphi\, e^{-i\xi} -  \eps^z_3\, s_{z0}\, s_{\varphi} - \eps^z_3\, c_\varphi\, e^{-i\xi}\right|}},\nn\\ \nn \\
s_z = \left|U_\eps\left(1,3\right)\right| &=& \left|s_{z0} + c_{z0}\,c_\varphi\,\eps^z_2 + c_{z0}\,s_\varphi\,\eps^z_3\right|.
\eea
An equation which allows to solve for the Dirac phase $\d$ is obtained, say, by equating the $(2,1)^{\mbox{\tiny th}}$ of $U_\eps$, after suitably rephased by $e^{-i\,\psi_2}$, with the corresponding one of $V^{*}_{\mbox{\tiny PMNS}}$ to get:
\bea
\label{delta}
 &&-\displaystyle{{1\over\sqrt{2}}}\,\left(s_{z0}\, c_\varphi + s_\varphi\, e^{-i\xi} - \eps_1^{z*}\, s_{z0}\, s_\varphi  + \eps_1^{z*}\, c_\varphi\, e^{-i\xi} + c_{z0}\, \eps_2^{z*}\right)\,e^{-i \phi_1}\, e^{-i \psi_2} = \nn\\
&& \left(-c_x\, s_y\, s_{z0} -s_x\, c_y \, e^{i \d}\right)\, e^{-i \rho}.
 \eea
Other equations, resulting from other entries, might be necessary to determine fully and consistently the Dirac phase.
\end{itemize}

\section{Realization}
\begin{itemize}
\item
Applying the similarity transformation Eq. (\ref{similarity}) on the ``non-rotated'' matrices of Eq. (\ref{fnon-rotated}), we get
\bea \label{tildef}
{\bf \tilde{f}}^{1}  =
\left(
\begin {array}{ccc}
A^z_1 &B^z_1& B^z_1\\
B^z_1& C^z_1& D^z_1\\
B^z_1&D^z_1&C^z_1
\end {array}\right),\;
 {\bf \tilde{f}}^{2}  =
\left(
\begin {array}{ccc}
A^z_2 &B^z_2& B^z_2\\
B^z_2& C^z_2& D^z_2\\
B^z_2&D^z_2&C^z_2
\end {array}\right),\;
{\bf \tilde{f}}^{3}  =
\left(
\begin {array}{ccc}
0 &B^z_3& -B^z_3\\
E^z_3 & C^z_3& D^z_3\\
-E^z_3&-D^z_3&-C^z_3
\end {array}
\right),
\eea
where $\left\{A^z_i, B^z_i, C^z_i, D^z_i\right\}$, $\left( i=1, 2\right)$, and $\left\{B^z_3, C^z_3, D^z_3, E^z_3\right\}$ are defined as
\bea
\label{tildefecoef}
&& A^z_i = c_z^2\, A_i + s_z^2\, \left(C_i + D_i\right),\;\;B^z_i = -\displaystyle{{t_{2z}\over 2\,\sqrt{2}}}\, \left(A^z_i - C^z_i -D^z_i\right), \nn \\ \nn \\
&&C^z_{i} = \displaystyle{{s_z^2\over 2}}\, A_i  + \displaystyle{{1\over 2}}\, \left(1+c_z^2\right)\, C_i -\displaystyle{{s_z^2\over 2}}\, D_i,\;\;
D^z_{i} = \displaystyle{{s_z^2\over 2}}\, A_i  + \displaystyle{{1\over 2}}\, \left(1+c_z^2\right)\, D_i -\displaystyle{{s_z^2\over 2}}\, C_i,\;\;\nonumber \\\nn\\
&&C^z_3 =  -\displaystyle{{t_{z}\over \sqrt{2}}}\,\left(B^z_3+E^z_3\right),\;
D^z_3 = \displaystyle{{t_{z}\over \sqrt{2}}}\,\left(B^z_3-E^z_3\right),\;
B^z_3 = c_z\, B_3,\; E_3^z = c_z\, E_3.
\eea

\item
Applying the similarity transformation Eq. (\ref{similarity}) on the ``non-rotated'' matrices of Eq. (\ref{gnon-rotated}), we get
\bea \label{tildeg}
{\bf \tilde{g}}^{1}  =
\left(
\begin {array}{ccc}
{\cal A}^z_1 &{\cal B}^z_1& {\cal B}^z_1\\
{\cal B}^z_1& {\cal C}^z_1& {\cal D}^z_1\\
{\cal B}^z_1&{\cal D}^z_1&{\cal C}^z_1
\end {array}\right),\;
{\bf \tilde{g}}^{(2)}  =
\left(
\begin {array}{ccc}
0 &{\cal B}^z_2& - {\cal B}^z_2\\
{\cal E}^z_2& {\cal C}^z_2& {\cal D}^z_2\\
-{\cal E}^z_2&- {\cal D}^z_2&-{\cal C}^z_2
\end {array}
\right),\;
{\bf \tilde{g}}^{3}  =
\left(
\begin {array}{ccc}
{2\cal A}^z_3 &{\cal B}^z_3& {\cal B}^z_3\\
{\cal E}^z_3& -{\cal A}^z_3& -{\cal A}^z_3\\
{\cal E}^z_3&-{\cal A}^z_3&-{\cal A}^z_3
\end {array}
\right)
\eea
where
\bea
\label{tildegecoef}
&&{\cal B}^z_1 =-\displaystyle{\frac{t_{2z}}{2\sqrt{2}}}\; \left({\cal A}^z_1 - {\cal C}^z_1 -{\cal D}^z_1\right),\;\;
{\cal  A}^z_1 = c_z^2\; {\cal A}_1 + s_z^2\; \left({\cal C}_1 + {\cal D}_1\right),\;\;
{\cal C}^z_{1} =\displaystyle{{s_z^2\over 2}}\; \left({\cal A}_1 - {\cal D}_1\right) + \displaystyle{{1\over 2}}\; \left(1+c_z^2\right)\; {\cal C}_1,  \nonumber \\
&& {\cal D}^z_{1} = \displaystyle{{s_z^2\over 2}}\; \left({\cal A}_1 - {\cal C}_1\right) + \displaystyle{{1\over 2}}\; \left(1+c_z^2\right)\; {\cal D}_1,\;\;
{\cal C}^z_2 = -\displaystyle{{\frac{t_{z}}{\sqrt{2}}}}\;\left({\cal B}^z_2 + {\cal E}^z_2 \right),\;\;
{\cal D}^z_2 = \displaystyle{\frac{t_{z}}{\sqrt{2}}}\;\left({\cal B}^z_2 - {\cal E}^z_2 \right),  \nonumber \\
&& {\cal B}^z_2 = c_z\;\ {\cal B}_2,\;\; {\cal E}^z_2 = c_z\; {\cal E}_2,\;\;
{\cal A}^z_3 = \displaystyle{{\frac{t_{2z}}{2\sqrt{2}}}}\; \left({\cal B}^z_3 + {\cal E}^z_3 \right),\;\;
{\cal B}^z_3 = c^2_z\; {\cal B}_3 - s^2_z\; {\cal E}_3,\;\;
{\cal E}^z_3 = c^2_z\; {\cal E}_3 - s^2_z\; {\cal B}_3
\eea

\item
Applying the similarity transformation Eq. (\ref{similarityh}) on the ``non-rotated'' matrices of Eq. (\ref{hnon-rotated}), we get the symmetric matrices
\bea
\label{tildeh}
{\bf \tilde{h}}^{1}  &=&
\left(
\begin {array}{ccc}
A^z_{R1} &-\displaystyle{\frac{t_z}{\sqrt{2}}}\; A^z_{R1} & -\displaystyle{\frac{t_z}{\sqrt{2}}}\; A^z_{R1}\\\\
\left(\tilde{h}^{1}\right)_{12}& \displaystyle{\frac{t^2_z}{2}}\; A^z_{R1}& \displaystyle{\frac{t^2_z}{2}}\; A^z_{R1}\\\\
\left(\tilde{h}^{1}\right)_{13}&\left(\tilde{h}^{1}\right)_{23}&
\displaystyle{\frac{t^2_z}{2}}\; A^z_{R1}
\end {array}
\right),
\nn\\\nn\\
{\bf \tilde{h}}^{2} & = &
\left(
\begin {array}{ccc}
t_z^2 \;\left(C^z_{R2} + D^z_{R2}\right) &\displaystyle{\frac{t_z}{\sqrt{2}}}\; \left(C^z_{R2} + D^z_{R2}\right) & \displaystyle{\frac{t_z}{\sqrt{2}}}\; \left(C^z_{R2} + D^z_{R2}\right)\\\\
\left(\tilde{h}^{2}\right)_{12}& C^z_{R2}& D^z_{R2}\\\\
\left(\tilde{h}^{2}\right)_{13}&\left(\tilde{h}^{2}\right)_{23}&C^z_{R2}
\end {array}
\right),
\eea
where
\bea
\label{tildehecoef}
A^z_{R1} = c^2_z\; A_{R1},\;\;
C^z_{R2} = \displaystyle{\frac{1}{2}}\; \left(1+c^2_z\right)\; C_{R2} -
\displaystyle{\frac{s_z^2}{2}}\;   D_{R2},\;\;
D^z_{R2} = -\displaystyle{\frac{s_z^2}{2}}\;  C_{R2} + \displaystyle{\frac{1}{2}}\; \left(1+c^2_z\right)\; D_{R2}.
\eea

\item
Applying the similarity transformation Eq. (\ref{similarityG}) on the ``non-rotated'' matrices of Eq. (\ref{Gnon-rotated}), we get
\bea
\label{tildeG}
{\bf \tilde{g}}^{1} & =&
\left(
\begin {array}{ccc}
{\cal A}^z_1 &-\displaystyle{\frac{t_z}{\sqrt{2}}}\; {\cal A}^z_1& -\displaystyle{\frac{t_z}{\sqrt{2}}}\; {\cal A}^z_1\\\\
-\displaystyle{\frac{t_z}{\sqrt{2}}}\; {\cal A}^z_1& \displaystyle{\frac{t_z^2}{2}}\;{\cal A}^z_1& \displaystyle{\frac{t_z^2}{2}}\;{\cal A}^z_1\\\\
-\displaystyle{\frac{t_z}{\sqrt{2}}}\; {\cal A}^z_1&\displaystyle{\frac{t_z^2}{2}}\;{\cal A}^z_1 & \displaystyle{\frac{t_z^2}{2}}\;{\cal A}^z_1
\end {array}
\right),\;\;
{\bf \tilde{g}}^{2}  =
\left(
\begin {array}{ccc}
0 &{\cal B}^z_2& - {\cal B}^z_2\\\\
0& -\displaystyle{\frac{t_z}{\sqrt{2}}}\; {\cal B}^z_2& \displaystyle{\frac{t_z}{\sqrt{2}}} {\cal B}^z_2\\\\
0&- \displaystyle{\frac{t_z}{\sqrt{2}}}\; {\cal B}^z_2&\displaystyle{\frac{t_z}{\sqrt{2}}}\; {\cal B}^z_2
\end {array}
\right), \nn\\\nn\\
{\bf \tilde{g}}^{3} & =&
\left(
\begin {array}{ccc}
t_z^2\; \left({\cal C}_3^z + {\cal D}_3^z\right) &\displaystyle{\frac{t_z}{\sqrt{2}}}\; \left({\cal C}_3^z + {\cal D}_3^z\right)& \displaystyle{\frac{t_z}{\sqrt{2}}}\; \left({\cal C}_3^z + {\cal D}_3^z\right)\\\\
\displaystyle{\frac{t_z}{\sqrt{2}}}\;\left({\cal C}_3^z + {\cal D}_3^z\right)& {\cal C}_3^z& {\cal D}_3^z\\\\
\displaystyle{\frac{t_z}{\sqrt{2}}}\; \left({\cal C}_3^z + {\cal D}_3^z\right)&
{\cal D}_3^z&{\cal C}_3^z
\end {array}
\right),\;\;
{\bf \tilde{g}}^{4}  =
\left(
\begin {array}{ccc}
\sqrt{2}\; t_z\; {\cal B}^z_4  &{\cal B}^z_4& {\cal B}^z_4\\\\
-t_z^2\; {\cal B}^z_4& -\displaystyle{\frac{t_z}{\sqrt{2}}}\;  {\cal B}^z_4 & -\displaystyle{\frac{t_z}{\sqrt{2}}}\; {\cal B}^z_4 \\\\
-t_z^2\; {\cal B}^z_4&-\displaystyle{\frac{t_z}{\sqrt{2}}}\; {\cal B}^z_4 &-\displaystyle{\frac{t_z}{\sqrt{2}}}\; {\cal B}^z_4
\end {array}
\right),\nn\\
\eea
where
\bea \label{tildeGecoef}
{\cal A}^z_1 &=& c^2_z\; {\cal A}_1,\;\;
{\cal B}^z_2 = c_z\; {\cal B}_2,\;\;
{\cal C}^z_3 = \displaystyle{\frac{1}{2}}\;\left(1+c^2_z\right)\; {\cal C}_3 - \displaystyle{\frac{s_z^2}{2}}\; {\cal D}_3,\nn\\
{\cal D}^z_3 &=& -\displaystyle{\frac{s_z^2}{2}}\; {\cal C}_3 + \displaystyle{\frac{1}{2}}\; \left(1+c^2_z\right)\;{\cal D}_3,\;\;
{\cal B}^z_4 = c^2_z\; {\cal B}_4.
\eea

\end{itemize}

\end{document}